\pdfoutput=1

\documentclass[11pt,twoside,a4paper,cmspaper,final,collab]{cms-tdr}
\def\svnVersion{147276}\def\cmsCernNo{2012-029}\def\cmsCernDate{\today}\def\cmsMessage{Submitted to Physics Letters B}

\begin{document}\cmsNoteHeader{SUS-11-021}

\hyphenation{had-ron-i-za-tion}
\hyphenation{cal-or-i-me-ter}
\hyphenation{de-vices}

\RCS$Revision: 143211 $
\RCS$HeadURL: svn+ssh://alverson@svn.cern.ch/reps/tdr2/papers/SUS-11-021/trunk/SUS-11-021.tex $
\RCS$Id: SUS-11-021.tex 143211 2012-08-10 13:47:39Z alverson $
\ifthenelse{\boolean{cms@external}}{\providecommand{\cmsLeft}{top}}{\providecommand{\cmsLeft}{left}}
\ifthenelse{\boolean{cms@external}}{\providecommand{\cmsRight}{bottom}}{\providecommand{\cmsRight}{right}}
\cmsNoteHeader{SUS-11-021} 
\title{Search for physics beyond the standard model in events with a Z boson, jets, and missing transverse energy in pp collisions at \texorpdfstring{$\sqrt{s}=7$~TeV}{sqrt(s) = 7 TeV}}

\newcommand{\zjets}{\ensuremath{\Z+\text{jets}}\xspace}
\newcommand{\gjets}{\ensuremath{\gamma+\text{jets}}}
\newcommand{\lumifinal}{{4.98\fbinv}}
\newcommand{\JZB}{\ensuremath{\mathrm{JZB}}\xspace}
\newcommand{\cls}  {\ensuremath{\mathrm{CL_S}}}
\newcommand{\njets}{\ensuremath{N_{\text{jets}}}}
\date{\today}

\abstract{
A search is presented for physics beyond the standard model (BSM) in events with a $\Z$ boson, jets, and missing transverse energy (\MET).
This signature is motivated by BSM physics scenarios, including supersymmetry.
The study is performed using a sample of proton-proton collision data collected at $\sqrt{s}=7$\TeV with the
CMS experiment at the LHC, corresponding to an integrated luminosity of 4.98\fbinv.
The contributions from the dominant standard model backgrounds are estimated from data using two complementary
strategies, the jet-Z balance technique and a method based on modeling \MET\ with data control samples.
In the absence of evidence for BSM physics, we set limits on the non-standard-model contributions to event yields
in the signal regions and interpret the results in the context of simplified model spectra.
Additional information is provided to facilitate tests of other BSM physics models.
}

\hypersetup{%
pdfauthor={CMS Collaboration},%
pdftitle={Search for physics beyond the standard model in events with a Z boson, jets, and missing transverse energy in pp collisions at sqrt(s) = 7 TeV},%
pdfsubject={CMS},%
pdfkeywords={CMS, physics, supersymmetry, SUSY}}

\maketitle 

\section{Introduction}\label{sec:introduction}
	
This paper describes a search for physics beyond the standard model (BSM) in proton-proton collisions at a center-of-mass energy of 7\TeV.
Results are reported from a data sample collected with the Compact Muon Solenoid (CMS) detector at the Large Hadron Collider (LHC)
at CERN corresponding
to an integrated luminosity of \lumifinal.
This search is part of a broad program of inclusive, signature-based searches for BSM physics at CMS, characterized
by the number and type of objects in the final state. Since it is not known a priori how the BSM physics will be manifest,
we perform searches in events containing jets and missing transverse energy (\MET)~\cite{CMS1,CMS4,CMS6},
single isolated leptons~\cite{CMS2},
pairs of opposite-sign~\cite{CMS5} and same-sign~\cite{CMS3} isolated leptons, photons~\cite{CMS7,CMS8}, etc.
Here we search for evidence of BSM physics in final states containing a \Z boson
that decays to a pair of oppositely-charged isolated electrons or muons.
Searches for BSM physics in events containing oppositely-charged leptons have also been performed by the ATLAS collaboration~\cite{ATLAS1,ATLAS2,scalartops}.

This strategy offers two advantages with respect to other searches.
First, the requirement of a leptonically-decaying \Z boson significantly suppresses large standard model (SM) backgrounds
including QCD multijet production, events containing \Z bosons decaying to a pair of invisible neutrinos, and events
containing leptonically-decaying \PW\ bosons, and hence provides a clean environment in which to search for BSM physics.
Second, final states with \Z bosons are predicted in many models of BSM physics, such as supersymmetry (SUSY)~\cite{SUSY1,SUSY2,SUSY3,SUSY4,Baer:1990in}.
For example, the production of a $\Z$~boson in the decay $\chiz_2 \rightarrow \chiz_1~\Z$, where $\chiz_1$ ($\chiz_2$)
is the lightest (second lightest) neutralino,
is a direct consequence of the gauge structure of SUSY, and can become a favored  channel
in regions of the SUSY parameter space where the neutralinos have a large Higgsino or neutral Wino
component~\cite{Matchev:1999ft,Ruderman:2011vv,GMSB}.
Our search is also motivated by the existence of cosmological cold dark matter~\cite{WMAP}, which could consist
of weakly-interacting massive particles~\cite{WIMPS} such as the lightest SUSY neutralino in R-parity conserving
SUSY models~\cite{RParity}. If produced in pp collisions, these particles would escape detection and yield events with large \MET.
Finally, we search for BSM physics in events containing hadronic jets. This is motivated by the fact that new,
heavy, strongly-interacting particles predicted by many BSM scenarios may be produced with a large cross section and hence be observable
in early LHC data, and such particles tend to decay to hadronic jets.
These considerations lead us to our target signature consisting of a leptonically-decaying \Z boson produced
in association with jets and \MET.

After selecting events with jets and a $\Z\to\ell^+\ell^-$ ($\ell=\Pe,\mu$) candidate,
the dominant background consists of SM \Z production accompanied by jets from initial-state radiation (\zjets).
The \MET\ in \zjets\ events arises primarily when jet energies are mismeasured.
The \zjets\ cross section is several orders of magnitude larger
than our signal, and the artificial \MET\ is not necessarily well reproduced in simulation.
Therefore, the critical prerequisite to a discovery of BSM physics in the $\Z+\mathrm{jets}+\MET$ final state is
to establish that a potential excess is not due to SM \zjets\ production accompanied by artificial
\MET\ from jet mismeasurements.
In this paper, we pursue two complementary strategies, denoted the Jet-\Z Balance (JZB) and \MET
template (MET) methods, which rely on different techniques to suppress the SM
\zjets\ contribution and estimate the remaining background.
The two methods employ different search regions, as well as
different requirements on the jet multiplicity and \Z boson identification.
After suppressing the \zjets\ contribution, the most
significant remaining SM background consists of  events with a pair of top quarks that both decay leptonically (dilepton \ttbar).
We  exploit the fact that in dilepton \ttbar events the two lepton flavors are uncorrelated,
which allows us to use a control sample of $\Pe\Pgm$ events, as well as events in the sideband
of the dilepton mass distribution, to estimate this background.

The \JZB method is sensitive to BSM models where the \Z boson and dark matter candidate
are the decay products of a heavier particle. In such models, the \Z boson and \MET directions are correlated,
with the strength of this correlation dependent on the BSM mass spectrum.
The \zjets\ background contribution to the \JZB signal region is estimated from a \zjets\ sample,
by exploiting the lack of correlation between the direction of the \Z boson and \MET\
in these events for large jet multiplicity.
With this method, the significance of an excess is reduced in models where the \MET and \Z directions
are not correlated.

The MET method relies on two data control samples, one consisting of events with photons accompanied by jets
from initial-state radiation (\gjets) and one consisting of
QCD multijet events, to evaluate the \zjets\ background in a high \MET signal region.
In contrast to the \JZB method, the MET method does not presume a particular mechanism
for the production of the \Z boson and \MET. The significance of an
excess is reduced in models that also lead to an excess in both the
$\text{jets}+\MET$ and $\gamma+\text{jets}+\MET$ final states.

The paper is organized as follows: we first describe the detector (Section~\ref{sec:detector}), and the data and simulated samples
and event selection that are common to both strategies (Section~\ref{sec:eventsel}). The two methods are then described
and the results presented (Sections~\ref{sec:jzb} and~\ref{sec:met}).
Systematic uncertainties on the signal acceptance and efficiency are presented in Section~\ref{sec:systematics}.
Next, the two sets of results are interpreted in the
context of simplified model spectra (SMS)~\cite{Alves:2011wf,ArkaniHamed:2007fw,SMS1},
which represent decay chains of new particles that may occur in a wide variety of BSM physics scenarios, including SUSY (Section~\ref{sec:interpret}).
We provide additional information to allow our results to be applied to arbitrary BSM physics scenarios (Section~\ref{sec:outreach}).
The results are summarized in Section~\ref{sec:conclusion}.

\section{The CMS Detector}\label{sec:detector}

The central feature of the CMS apparatus is a superconducting solenoid
of 6\unit{m} internal diameter, providing a field of 3.8\unit{T}. Within the field volume are the silicon pixel
and strip tracker, the crystal electromagnetic calorimeter, and the brass/scintillator
hadron calorimeter. Muons are measured in gas-ionization detectors embedded in the steel return yoke.
In addition to the barrel and endcap detectors, CMS has extensive forward calorimetry.
The CMS coordinate system is defined with the origin at the center of the detector and the
$z$ axis along the direction of the counterclockwise beam. The transverse plane is perpendicular to the beam axis,
with $\phi$ the azimuthal angle, $\theta$ the polar angle, and $\eta=-\ln[\tan(\theta/2)]$
the pseudorapidity.
Muons are measured in the range $|\eta|< 2.4$.
The inner tracker measures charged particles within the range $|\eta| < 2.5$.
A  more detailed description of the CMS detector can be found elsewhere~\cite{JINST}.

\section{Samples and Event Selection}\label{sec:eventsel}

Events are required to satisfy at least one of a set of $\Pe\Pe$, $\Pe\Pgm$ or $\Pgm\Pgm$
double-lepton triggers, with lepton transverse momentum (\pt) thresholds of 17\GeV for one lepton and 8\GeV for the other.
Events with two oppositely-charged leptons
($\EE$, $\Pe^{\pm}\Pgm^{\mp}$, or $\MM$) are selected.
Details of the lepton reconstruction and identification can be found in Ref.~\cite{ref:electrons} for electrons and in Ref.~\cite{ref:muons} for muons.
Both leptons must have
$\pt > 20\GeV$, in the efficiency plateau of the triggers.
Electrons (muons) are restricted to $|\eta| < 2.5$ (2.4).
For the candidate sample, only $\EE$ and $\MM$ events are used, and the dilepton system is required
to have an invariant mass consistent with the mass of the \Z boson ($m_{\Z}$).
The $\Pe\mu$ events are used as a data control sample to estimate the $t\bar{t}$ background.

Because leptons produced in the  decays of low-mass particles, such as
hadrons containing \cPqb\  and \cPqc\ quarks,  are  nearly  always inside  jets,  they can  be
suppressed by requiring the leptons to be isolated in space from other
particles that carry a  substantial amount of transverse momentum.
The lepton isolation~\cite{ref:top} is defined using the
scalar sum of both the transverse momentum depositions in the calorimeters and the transverse momenta of tracks in a cone of
$\Delta{}R\equiv\sqrt{(\Delta\eta)^2+(\Delta\phi)^2}<0.3$
around each lepton, excluding the lepton itself.
Requiring the ratio of this sum to the lepton \pt\ to be smaller than 15\% rejects
the large background arising from QCD production of jets.

We select jets~\cite{ref:jes} with  $\pt > 30\GeV$ and  $|\eta| < 3.0$, separated  by $\Delta  R  > 0.4$ from  leptons  passing the  analysis
selection.
We use the particle flow (PF) method~\cite{CMS-PAS-PFT-10-002} to reconstruct charged and neutral hadrons,
muons, electrons, and photons.
The PF objects are clustered to form jets using the anti-$k_{\mathrm{T}}$ clustering algorithm~\cite{antikt}
with a distance parameter of 0.5, as implemented in the {\sc fastjet} package~\cite{Cacciari:2005hq,FastJet}.
We apply \pt- and $\eta$-dependent corrections to account for residual effects of non-uniform detector response.
The contribution to the jet energy from pile-up is estimated on an event-by-event basis using the
jet area method described in Ref.~\cite{cacciari-2008-659}, and is subtracted from the overall jet \pt.
The missing transverse momentum \MET is defined as the magnitude of the vector sum of the transverse momenta of all PF objects.
The \MET vector is the negative of that same vector sum.

The sample passing the above preselection requirements is dominated by SM \zjets\ events,
which must be suppressed in order to achieve sensitivity to BSM physics.
As discussed in the introduction, we pursue two complementary approaches to evaluate the
\zjets\ background. 
Samples of \zjets, \ttbar, $\PW\PW$, $\PW\Z$, and $\Z\Z$  Monte Carlo (MC) simulated events generated with
\MADGRAPH5.1.1.0~\cite{Alwall:2007st} are used to guide the design of these methods, but the dominant
backgrounds are estimated with techniques based on data control samples.
Events produced by \MADGRAPH are passed to \PYTHIA6.4.22~\cite{Sjostrand:2006za}
for the generation of parton showers.
Additional MC samples of \zjets, \gjets, and QCD multijet events generated with
\PYTHIA6.4.22 are used to validate the \MET\ template method of Sec.~\ref{sec:met}.
We also present the expected event yields for two benchmark scenarios of the constrained minimal
supersymmetric extension of the standard model (CMSSM)~\cite{CMSSM}, denoted LM4 and LM8~\cite{PTDR2},
which are generated with the same version of \PYTHIA.
The CMSSM is described with five parameters: the universal scalar and gaugino masses $m_0$ and $m_{1/2}$,
the universal soft SUSY-breaking parameter $A_0$, the ratio of vacuum expectation values of the two Higgs doublets
$\tan\beta$, and the sign of the Higgs mixing parameter $\mu$.
The LM4 (LM8) parameter sets are $m_{0} = 210\ (500)\GeV$, $m_{1/2}=285\ (300)\GeV$, $\tan\beta=10$, sign($\mu) = +$,
and $A_{0}=0\ (-300)\GeV$. The LM4 scenario is excluded in Ref.~\cite{CMS6}; this paper is the first to exclude LM8.
In these two scenarios heavy neutralinos predominantly decay to a $\Z$ boson and a lighter neutralino.
All samples are generated using the CTEQ6~\cite{Pumplin:2002vw} parton distribution functions (PDFs)
and normalized to next-to-leading order (NLO) cross sections.
Simulation of the CMS detector response is performed using
\GEANTfour~\cite{Agostinelli:2002hh}.
The simulated events are subsequently reconstructed and analyzed in the same way as the data,
and are rescaled to describe the measured distribution of overlapping pp collisions
in the same bunch crossing (referred to as ``pile-up reweighting'').

\section{\texorpdfstring{\JZB}{JZB} Search}\label{sec:jzb}

\subsection{Jet-\texorpdfstring{\Z}{Z} Balance Variable}

The \JZB variable is defined in the $xy$ plane as

\begin{equation}
\JZB =  \Big\lvert\sum_{\text{jets}}\vec{\pt}\Big\rvert      - \Big\lvert\vec{\pt}^{(\Z)}\Big\rvert %
     \approx  \left|-\VEtmiss-\vec{\pt}^{(\Z)}\right| - \Big\lvert\vec{\pt}^{(\Z)}\Big\rvert.
\end{equation}

Thus \JZB measures the imbalance between the \pt of the \Z boson and that of the hadronic system.
In SM \zjets\ events, the \JZB distribution is approximately symmetric about zero, while for BSM
physics it may be asymmetric, due to correlated production of the \Z boson and invisible particles.
Five signal regions are defined by requirements
on the \JZB event variable, from $\JZB>50\GeV$ to $\JZB>250\GeV$ in steps of 50\GeV.
The signal region in the invariant mass distribution is defined as $|m_{\ell\ell}-m_{\Z}|<20\GeV$.

In SM \zjets events, the \JZB variable is analogous to \MET with sign information.
The sign depends on whether \MET is due to an under- or over-measurement of the jet energy.
The probability of a downward fluctuation of the jet energy measurement is in general higher than the
probability of an upward fluctuation, leading to an asymmetry of the JZB distribution in SM $Z$ events
with exactly 1 jet. However, the \JZB distribution in SM \zjets events becomes more Gaussian with increasing
jet multiplicity, because in multijet events the direction of a mismeasured jet is uncorrelated with the
direction of the \Z boson.
Already in three-jet events, where in the most probable configuration, the two leading jets are
back-to-back~\cite{Berger}, instrumental effects largely cancel.
For this reason the \JZB method focuses on events containing at least three jets.

We search for BSM events where the \Z boson is the decay product of a heavier (parent) particle of mass $m_{M}$
and is produced in conjunction with an undetectable decay product of mass $m_\mathrm{X}$, which gives rise to \MET.
Let $p^{*}$ be the characteristic momentum of the decay products in the rest frame of the parent particle.
If the parent particle has a mass of the order of the electroweak scale, $m_{M} \sim O(m_\mathrm{X} + m_{\Z})$,
$p^{*}$ is small, and $p^{*}$ can be smaller than the laboratory momentum of the parent.
In that case, the daughter particles all appear in a tightly collimated angular region, the transverse momenta
of the \Z and invisible particle are balanced by the other particles in the decay chain, and large values of \JZB can ensue.
An example of such a decay chain is
$\PSg \rightarrow  \cPaq +  \PSq \rightarrow \cPaq + \cPq +\chiz_2 \rightarrow \cPaq + \cPq + \Z + \chiz_1$,
where $\PSg$, $\PSq$, and $\chiz_\mathrm{1,2}$ are the gluino, squark, and neutralino supersymmetric particles.

The signal and background discrimination arising from the angular correlation between the \Z boson and \MET\ can
be reduced in certain circumstances. For example, in $R$-parity-conserving SUSY, supersymmetric particles are produced
in pairs and there are two decay chains with one undetected lightest stable particle (LSP) at the end of each chain.
It can happen that the two unobserved particle momenta cancel each other, leading to small \MET and \JZB values.
Such configurations are, however, disfavored by the selection of events with significant \MET,
or large \JZB, which is equivalent to requiring that the two LSPs do not balance.
The angular correlation is therefore preserved in events with significant \MET.

To summarize, the balance between the jet system and the $\Z + \MET$ system leads to large, positive \JZB in events
where \MET and the \Z boson are pair-produced, while the $\JZB>0$ and $\JZB<0$ regions are evenly populated
in SM \zjets events.

\subsection{Background Determination}

The principal SM backgrounds are divided in two categories.
Backgrounds that produce opposite-flavor (OF) pairs $(\Pep\Pgmm, \Pem\Pgmp)$ as often as same-flavor (SF)
pairs $(\Pep\Pem, \Pgmp\Pgmm)$ are referred to as ``flavor--symmetric backgrounds''.
This category is dominated by \ttbar processes. Backgrounds with two SF leptons
from a \Z boson are referred to as ``\Z boson backgrounds''. This category is dominated by SM \zjets\ production.

Three non-overlapping data control regions are used to predict the contribution of flavor-symmetric backgrounds:
(a) OF events compatible with the \Z boson mass hypothesis (referred to as ``\Z-peak region''),
(b) OF events in the sideband of the \Z boson mass peak, and (c) SF events in this sideband.
The sideband region is defined as the union of $55<m_{\ell\ell}<70\GeV$ and $112<m_{\ell\ell}<160\GeV$;
it is chosen so that it includes the same number of events as the \Z-peak region in \ttbar simulation.
The two OF data control samples are compared in the region $30\GeV<|\JZB|<50\GeV$, which is outside the signal regions
and has little contribution from signal or $\Z(\to\tau\tau)+\text{jets}$.
The event yields from the two data control samples in this region are found to be in good agreement with each other
and with expectations from the MC simulation.
The systematic uncertainties on the number of events estimated from the three data control regions are assessed using
a large sample of simulated \ttbar events. The \JZB\ distribution in the SF \Z-peak (signal) region is found to agree
well with the corresponding distributions in the three control regions.
A 25\% uncertainty is assigned to each individual estimate in order
to cover discrepancies at large \JZB values, where the number of MC events is low, as well as small
differences between the data and MC simulation in the shape of the \JZB distribution.

The total contribution from flavor-symmetric backgrounds in the signal region is computed as the average of
the yields in the three data control regions, as they provide independent estimates of the same background process.
The systematic uncertainties assigned to these yields are approximately uncorrelated, and hence are added quadratically.
The absence of strong correlation is confirmed in MC simulation, as well as from the aforementioned comparison
of the number of events in the $30\GeV<|\JZB|<50\GeV$ region.

SM backgrounds with a reconstructed \Z boson are estimated using the negative \JZB region after subtraction of
flavor-symmetric backgrounds. This procedure relies on the fact that \zjets events with three or more jets evenly populate the negative
and positive sides of the \JZB distribution, as described above. The method is validated using
a large sample of simulated \zjets\ events and the \JZB distributions in the negative and positive \JZB regions
are found to agree very well. We assign a 25\% systematic uncertainty to the corresponding
prediction in order to cover small differences between the data and MC simulation in the shape of the \JZB distribution.

Other backgrounds, though less significant, are also accounted for in these estimates.
Contributions from the SM $\PW\Z$ and $\Z\Z$ processes are incorporated into the \zjets\ estimate,
since in these events the \MET and the \Z boson candidates do not share the same parent particle.
The background estimate from OF pairs accounts for $\PW\PW$, $\Z\rightarrow\tau\tau$, and single-top production.
Finally, events with one or more jets reconstructed as electrons or non-isolated leptons (from QCD multijet,
$\gamma+\textrm{jets}$, or electroweak processes) are accounted for by the background
estimate from the sideband control regions.

\begin{figure}[htbp]
 \begin{center}
  \includegraphics[width=0.45\textwidth]{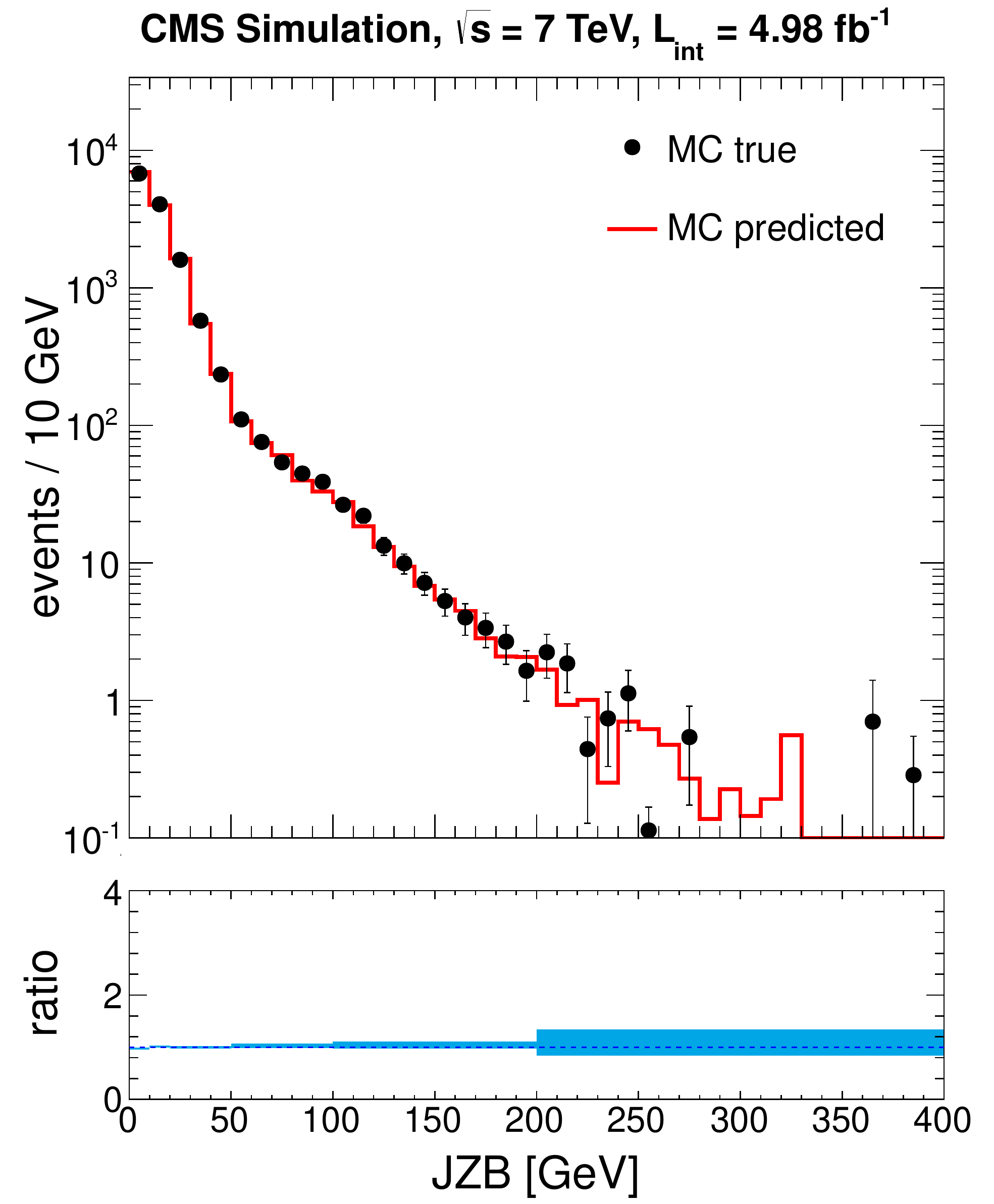}
  \includegraphics[width=0.45\textwidth]{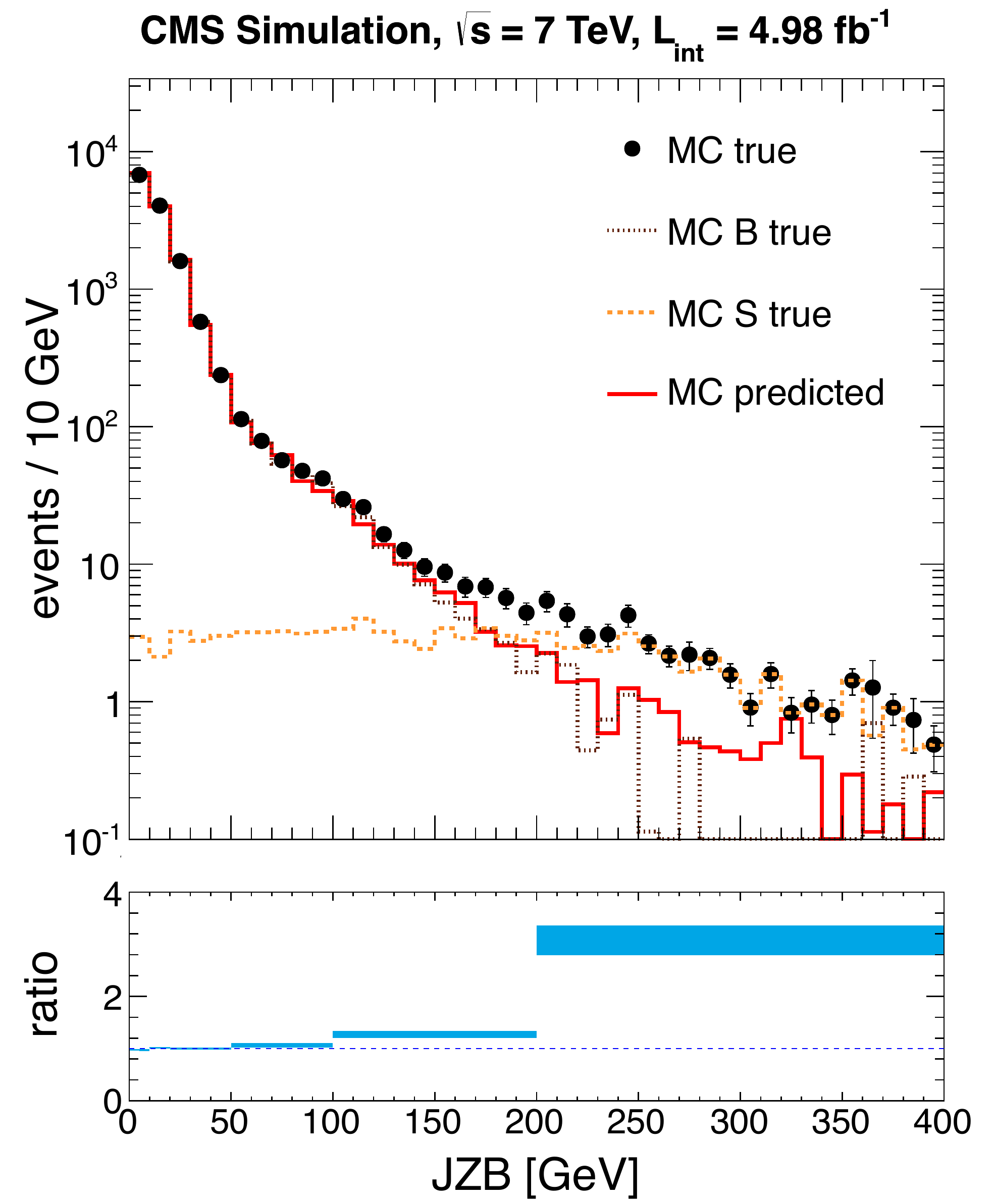}
  \caption{Comparison between true and predicted \JZB distributions in simulated samples for the
    background-only (\cmsLeft) and LM4-plus-background (\cmsRight) hypotheses.
    ``MC B'' and ``MC S'' denote the background and signal contributions to the true distribution, respectively.
    The lower plots show the ratio between true and predicted distributions.
    The error bars on the true distribution and in the ratio indicate the statistical uncertainty only.}\label{fig:mcclosure}
  \end{center}
\end{figure}

The overall background prediction method is validated using a simulated sample including
all SM backgrounds, with and without the inclusion of LM4 signal events.
The comparison between the true and predicted distributions is shown in Fig.~\ref{fig:mcclosure} for the two cases.
The inclusion of LM4 signal slightly modifies the predicted distribution because of contribution from the signal
to the control regions.
The slope change around $\JZB=50\GeV$ corresponds to the region where the \ttbar\ background starts to dominate.
The integrated event yields for the various signal regions are summarized in Table~\ref{tab:mcclosure}.
We find that there is good agreement in the background-only case, while good sensitivity to a possible signal remains.

\begin{table}[htbp]
 \begin{center}
  \renewcommand{\arraystretch}{1.3}
  \topcaption{Comparison between true and predicted \JZB event yields in SM MC simulation for the various signal regions.
    Uncertainties on the true MC yields reflect the limited MC statistics.
    The first (second) uncertainty in the MC predicted yields indicates the statistical (systematic) component.
}\label{tab:mcclosure}
  \begin{tabular}{l c c}
   \hline
   \hline
   Region        & MC true & MC predicted\\
   \hline
  $\JZB>50\GeV$  & $420 \pm 11$     & $414  \pm16  \pm 59$ \\
  $\JZB>100\GeV$ & $102 \pm  5$     & $ 98  \pm 6  \pm 14$ \\
  $\JZB>150\GeV$ & $ 25 \pm  2.6$   & $ 24  \pm 3.4\pm  3.0$ \\
  $\JZB>200\GeV$ & $ 8.5\pm  1.6$   & $  7.8\pm 1.8\pm  1.1$ \\
  $\JZB>250\GeV$ & $ 2.2\pm  0.9$   & $  3.2\pm 1.2\pm  0.5$ \\
   \hline
   \hline
  \end{tabular}
  \end{center}
\end{table}

\subsection{Results}\label{ssec:jzbresults}

\begin{figure}[htb]
  \begin{center}
    \includegraphics[width=0.45\textwidth]{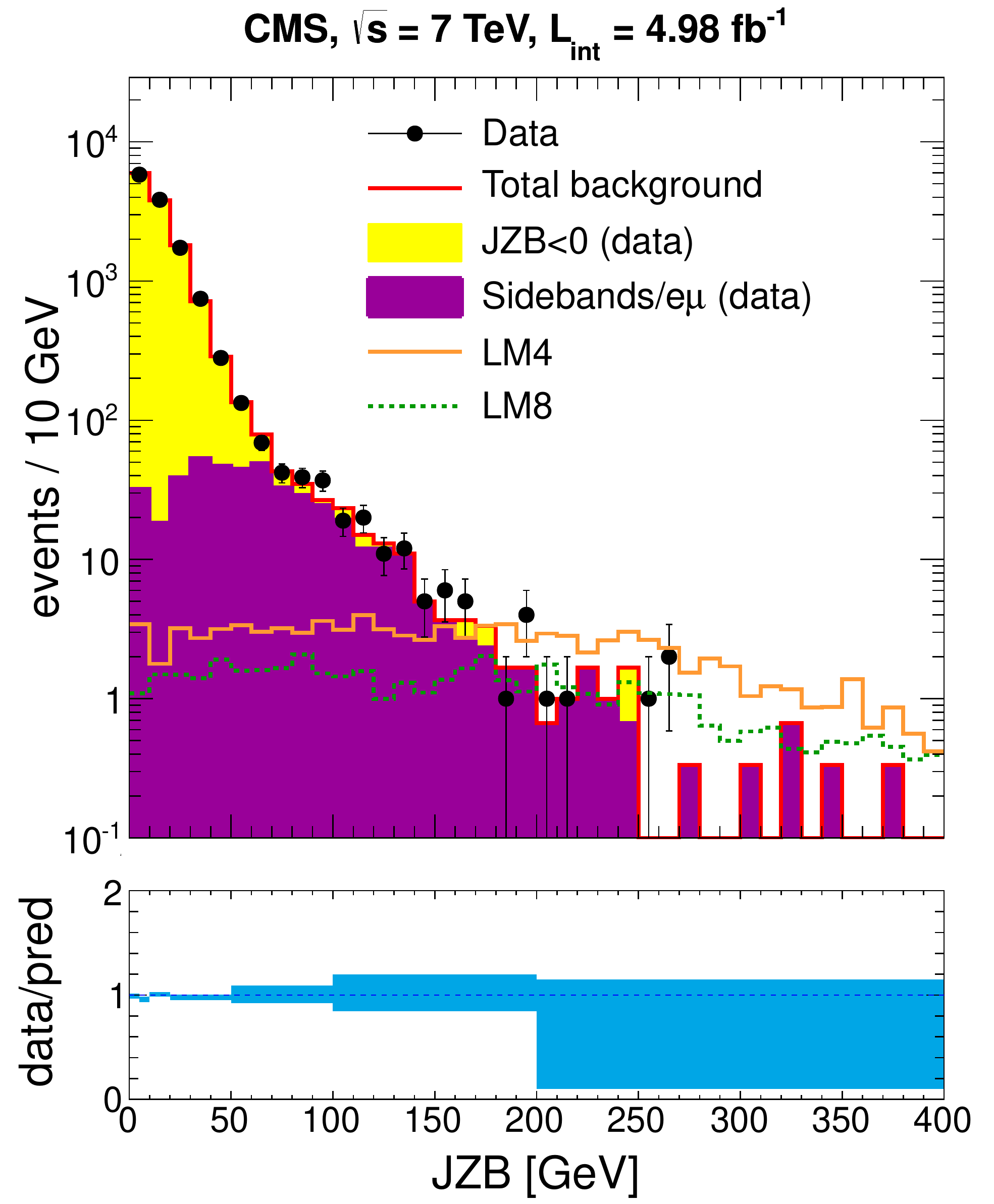}
    \caption{Comparison between the measured \JZB\ distribution in the $\JZB>0$ region
    and that predicted from data control samples. The distribution from the LM4 MC is overlaid.
    The bottom plot shows the ratio between the observed and predicted distributions.
    The error bars indicate the statistical uncertainties in data only.}
    \label{fig:BpredJZB}
  \end{center}
\end{figure}

The comparison between the observed and predicted distributions is shown in Fig.~\ref{fig:BpredJZB}.
The observed and predicted yields in the signal regions are summarized in Table~\ref{tab:results}, along
with 95\% confidence level (CL) upper limits on the yields of any non-SM process.
Upper limits are computed throughout this paper using a modified frequentist method (\cls)~\cite{CLS1,CLS2}.
The nuisance parameters (described in Section~\ref{sec:systematics}) are modeled with a lognormal distribution.
Table~\ref{tab:results} also shows the LM4 and LM8 yields, determined using NLO production cross sections.
These yields are corrected to account for the contribution of signal to the background control regions,
which tends to suppress the apparent yield of signal in the signal region.
The correction is performed by subjecting the signal samples to the same procedures as the data
and subtracting the resulting prediction from the signal yield in the signal region.
The expected	 LM4 and LM8 yields exceed the upper limits on the non-SM contributions to the yields
in the high \JZB signal regions.

\begin{table*}[hbtp]
\renewcommand{\arraystretch}{1.3}
\small
  \begin{center}
  \topcaption{Total number of events observed in the \JZB signal regions and corresponding background predictions
  from data control regions.
  The first uncertainty is statistical and the second systematic.
  For the observed yields, the first (second) number in parentheses is the yield in the \Pep\Pem (\Pgmp\Pgmm) final state.
  The 95\% CL upper limit (UL) on non-SM yields and the NLO yields for the LM4 and LM8 benchmark SUSY scenarios are also given,
  including the systematic uncertainties and the correction for signal contribution to the background control regions
  (see text for details).
  }\label{tab:results}
\begin{tabular}{l ccccc}
\hline
\hline
                 &  $\JZB>50\GeV$         &  $100\GeV$        &      $150\GeV$          &   $200\GeV$          & $250\GeV$      \\
\hline
\Z bkg           & $97\pm13\pm38$         & $8\pm3\pm3$       &     $2.7\pm1.8\pm0.8$   &    $1.0\pm1.0\pm0.3$ &  0            \\
Flavor-symmetric & $311\pm10\pm45$        & $81\pm5\pm12$     &     $19\pm3\pm3$        &    $7\pm2\pm1$       & $2.0\pm0.8\pm0.3$   \\
\hline
\hline
Total bkg        & $408\pm16\pm59$        & $89\pm6\pm12$     &     $22\pm3\pm3$        &    $8\pm2\pm1$       & $2.0\pm0.8\pm0.3$   \\
Data             & 408 (203,205)          & 88 (52,36)        &      21 (13,8)          &    5 (3,2)           &  3 (2,1)         \\
\hline
\hline
Observed UL      &   114                  &        32         &          14             &       6              &       6               \\
Expected UL      &   111                  &        31         &          13             &       7              &       4               \\
LM4              &  $62\pm4$              &    $52\pm4$       &       $40\pm4$          &    $29\pm4$          &     $18\pm4$            \\
LM8              &  $23\pm2$              &    $19\pm2$       &       $16\pm2$          &    $11.4\pm1.7$      &     $7.8\pm1.5$         \\
\hline
\hline
\end{tabular}
  \end{center}
\end{table*}

\section{MET Search}\label{sec:met}

For the MET method, we select events with two or more jets. Compared to the JZB method, the dilepton mass requirement is
tightened to $|m_{\ell\ell}-m_{\Z}|<10$\GeV, in order to further constrain mismeasurements of the lepton \pt's and to suppress
the \ttbar background.
As in the JZB method, the principal background is \zjets\ events. To suppress
this background, we require the events to have large \MET. Specifically, we define three signal regions:

\begin{itemize}
\item \MET $>$ 100\GeV (loose signal region);
\item \MET $>$ 200\GeV (medium signal region);
\item \MET $>$ 300\GeV (tight signal region).
\end{itemize}

The use of multiple signal regions allows us to be sensitive to BSM physics with differing \MET\ distributions.
To estimate the residual \zjets\ background with \MET\ from jet mismeasurements,
we model the \MET\ in \zjets\ events using $\gamma+\rm{jets}$ and QCD control samples in data.
After applying the \MET\ requirement, the dominant background is expected to be \ttbar in all three signal regions.
This background is estimated from a control sample of $\Pe\Pgm$ events in data.
Additional sub-leading backgrounds from $\PW\Z$ and $\Z\Z$ diboson production are estimated from simulation.

\subsection{Background Estimates}\label{sec:met_bkgestimate}

\subsubsection{\texorpdfstring{\zjets}{Z+jets} Background Estimate}
The background from SM \zjets\ production
is estimated using a \MET template method~\cite{Pavlunin:2009sx}.
In \zjets\ events, the \MET\ is dominated by mismeasurements of the hadronic system.
Therefore, the \MET distribution in these events can be modeled using a control sample
with no true \MET and a similar hadronic system as in \zjets\ events.
We use two complementary control samples: one consisting of \gjets\ events and one consisting
of QCD multijet events.
The \gjets\ (QCD multijet) events are selected with a set of single photon (single jet) triggers with online \pt\ tresholds varying
from 20--90~GeV (30--370~GeV).
To account for kinematic differences between the hadronic systems in the control and signal samples,
the expected \MET distribution of a \zjets\ event is obtained from the \MET distribution
of \gjets\ or QCD multijet events of the same jet multiplicity and scalar sum
of jet transverse energies, normalized to unit area; these normalized distributions are referred to as \MET\ templates.
The two control samples are complementary.
The  \gjets\ events have a topology
that is similar to the \zjets\ events, since both consist of a well-measured
object recoiling against a system of hadronic jets.
When selecting photons, we include hadronic jets
in which a large fraction of the energy is carried by photons or neutral pions.
Such jets are well measured; the \MET\ in these events arises from jets with a large hadronic energy fraction as in the true \gjets\ events.
The QCD multijet sample has better statistical precision due to the larger number of events, and eliminates
possible contributions to \MET\ from mismeasurement of the photon in the \gjets\ sample.
The \MET\ templates extracted from the QCD sample must be corrected for a
small bias of the \MET, which is observed in \gjets\ and \zjets\ events in the
direction of the recoiling hadronic system, due to a small systematic
under-measurement of the jet energies. This bias of the \MET\ is measured to be
approximately 6\% of the \pt\ of the hadronic recoil system, and the correction primarily affects the bulk of the \MET\ distribution.
A similar effect is present when using the \gjets\ templates because a minimum \pt\ threshold is applied to the photons but not to the \Z bosons.
However, the maximum resulting bias in the \MET\ is approximately 1\GeV, and is hence negligible.

Because jets in QCD dijet events have a different topology than those in $\Z+2$ jet events, the \gjets\ method alone is used to determine the \zjets\ background for events
with exactly two jets.  For events with at least three jets, we use the average of the background estimates from the \gjets\ and QCD multijets methods.
The two methods yield consistent predictions for events with at least three jets, which illustrates the robustness of the \MET\ template method and provides a cross-check
of the data-driven background prediction.
For the benchmark SUSY scenarios LM4 and LM8, we have verified that the impact
of signal contamination on the predicted background from the \MET template method is negligible.

The systematic uncertainty in the background prediction from the \gjets\ method is dominated by possible differences
between the predicted and true number of events when we apply the background estimate to the MC,
which is limited by the statistical precision of the MC samples (MC closure test, 30\% uncertainty).
Additional uncertainties are evaluated by varying the photon selection criteria (10\% uncertainty) and
from the difference in the number of reconstructed pile-up interactions in the \zjets\ and \gjets\ samples (5\% uncertainty).
The total uncertainty is 32\%.  The corresponding uncertainty in the background prediction from the QCD multijet method is
dominated by possible differences between the  predicted and true number of events in the MC closure test (ranging from 20\%
for \MET\ $>$ 30\GeV to 100\% for \MET $>$ 100\GeV). The uncertainty in the bias of the \MET\ in the direction of the hadronic
recoil contributes an additional 16\% uncertainty to this background prediction.

\subsubsection{Opposite-Flavor Background Estimate}

As in the JZB method, the \ttbar contribution is estimated using an OF subtraction technique,
based on the equality of the \ttbar\ yield in the OF and SF  final states after correcting
for the differences in the $\Pe$ and $\Pgm$ selection efficiencies.
Other backgrounds for which the lepton flavors are
uncorrelated (for example, $\PWp\PWm$, $\gamma^{*}/\Z\rightarrow \tau^+\tau^-$ and single-top processes, which are
dominated by the $\cPqt\PW$ production mechanism) are also included in this estimate.

To predict the SF yield in the \MET signal regions, we use the OF yield satisfying the same \MET requirements.
This yield is corrected using the ratio of
selection efficiencies $R_{\Pgm\Pe}\equiv\varepsilon_\mu/\varepsilon_e=1.07 \pm 0.07$,
which is evaluated from studies of $\Z\to\MM$ and $\Z\to\EE$
events in data. The uncertainty on this quantity takes into account a small variation with
respect to lepton \pt. To improve the statistical precision
of the background estimate, we do not require the OF events to lie in the $\Z$ mass region,
and we apply a scale factor $K=0.16 \pm 0.01$, extracted from simulation, to account for the fraction of \ttbar\ events
that satisfy $|m_{\ell\ell}-m_{\Z}|<10$\GeV.
The uncertainty in $K$ is determined by the difference between this quantity evaluated in data versus simulation.
An alternate method is to use OF events in the \Z mass window; scaling is not required, but fewer events are available.
This method yields a prediction that is consistent with that from the nominal method but with a larger statistical uncertainty.
The systematic uncertainty on the OF background prediction is dominated by a 25\% uncertainty in the yield predicted for the $\MET>200$\GeV region,
due to possible differences between the true and predicted number of events in MC closure tests.
The uncertainties in the correction factors $R_{\Pgm\Pe}$ (7\%) and $K$ (6\%) also contribute.

\subsubsection{Other Backgrounds}

Backgrounds from pairs of $\PW\Z$ and $\Z\Z$ vector bosons are estimated from MC, and a 50\% systematic
uncertainty is assessed based on comparison of simulation to data in events with jets and exactly 3 leptons ($\PW\Z$ control sample, MC expected purity approximately 90\%)
and exactly 4 leptons ($\Z\Z$ control sample, MC expected purity approximately 100\%), which have limited statistical precision due to small event yields.
Backgrounds from events with misidentified leptons are negligible due to the requirement of two isolated leptons with \pt\ $>$ 20\GeV in the \Z mass window.

\subsection{Results}
\label{sec:met_results}

\begin{figure}[thbp]
\begin{center}
\includegraphics[width=0.45\textwidth]{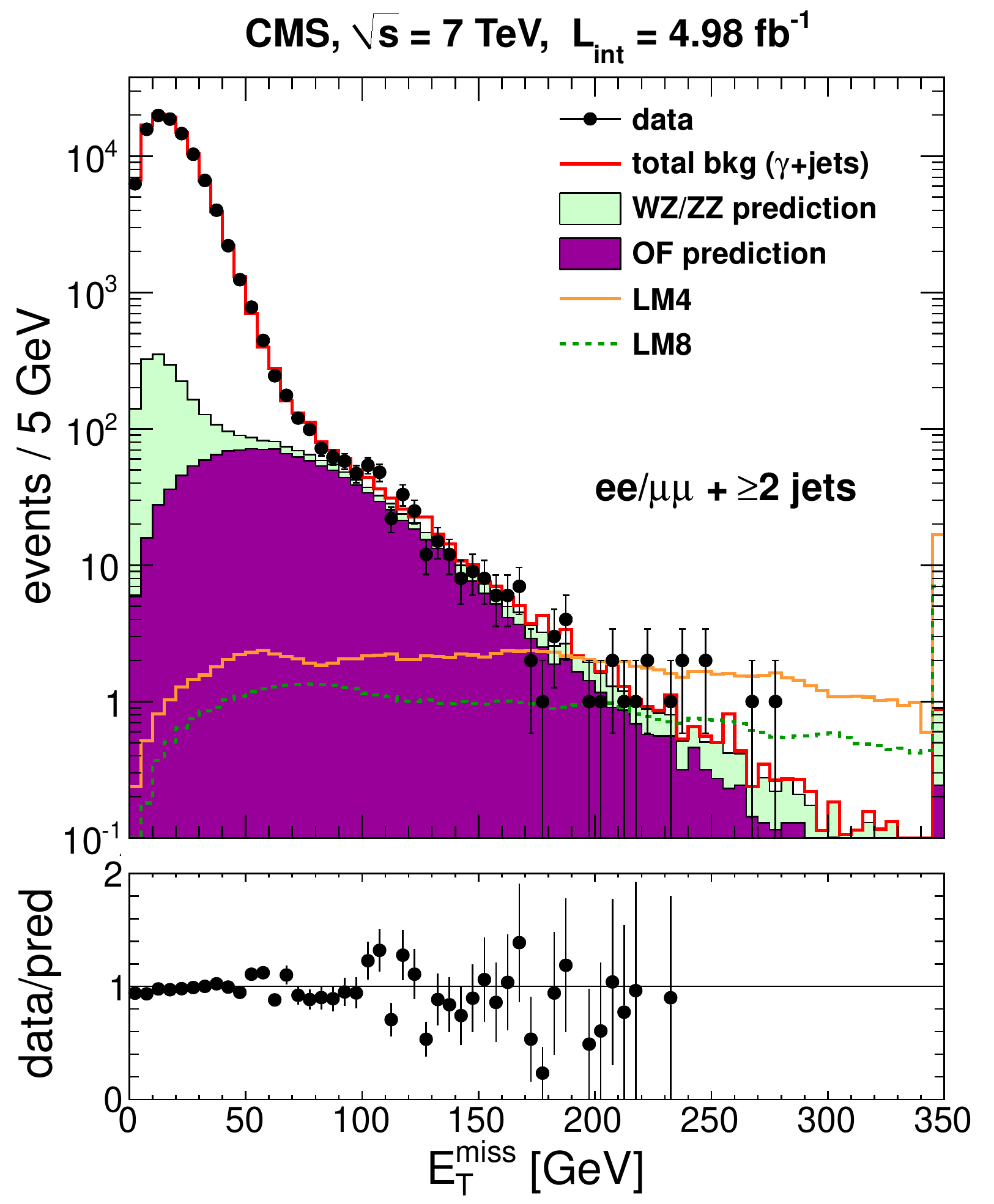}
\includegraphics[width=0.45\textwidth]{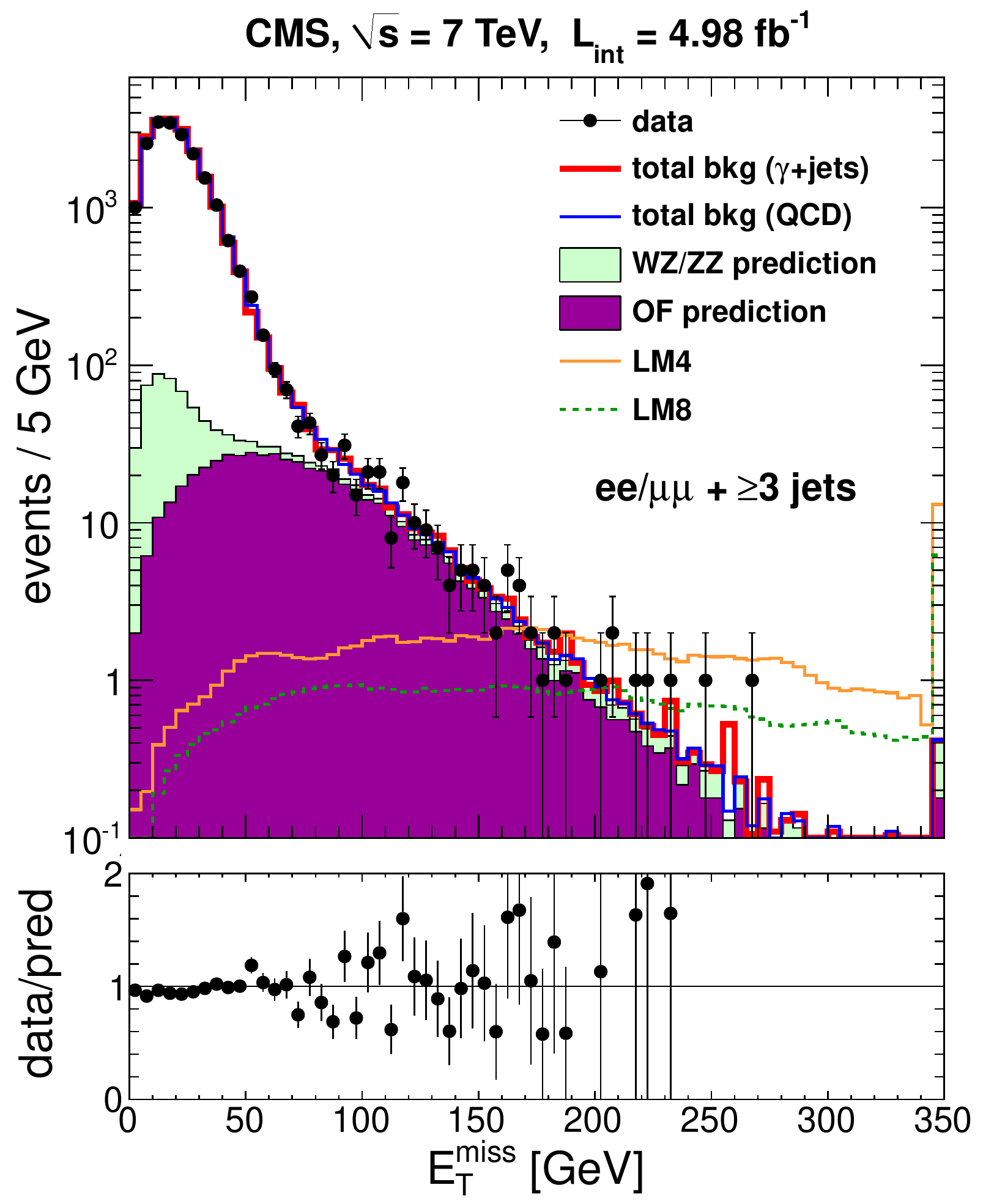}
\caption{\label{fig:results}\protect
The observed \MET distribution for events with \njets\ $\ge$ 2 (\cmsLeft) and \njets\ $\ge$ 3 (\cmsRight) for data (black points),
predicted OF background from simulation normalized to the $\Pe\mu$ yield in data (solid dark purple histogram),
$\PW\Z+\Z\Z$ background (solid light green histogram),
and total background including the \zjets\ predicted from \gjets\ (red line) and QCD (blue line) \MET\ templates.
The ratio of the observed and total predicted yields (data/pred) is indicated in the bottom plots using the \gjets\ (\cmsLeft) and average
of the \gjets\ and QCD (\cmsRight) methods.  The error bars indicate the statistical uncertainties in data only.
}
\end{center}
\end{figure}

The data and SM predictions are shown in Fig.~\ref{fig:results} and summarized in Table~\ref{resultsyieldtable}
(\njets\ $\ge$ 2) and Table~\ref{resultsyieldtable3} (\njets\ $\ge$ 3).
In addition to the loose, medium, and tight signal regions defined above, we quote the
predicted and observed event yields in two low \MET regions, which allows us to validate our background estimates
with increased statistical precision.
For all five regions, the
observed yields are consistent with the predicted background yields. No evidence for
BSM physics is observed. We place 95\% CL upper limits on the non-SM contributions to the yields in
the signal regions. These model-independent upper limits may be used
in conjunction with the signal efficiency model discussed in Section~\ref{sec:outreach}
to perform exclusions in the context of an arbitrary BSM physics model.
We quote results separately for \njets\ $\ge$ 2 and \njets\ $\ge$ 3 to improve the sensitivity to BSM models with low
and high average jet multiplicities, respectively.
We also quote the NLO expected yields for the SUSY benchmark processes LM4 and LM8, including the statistical component
and the systematic uncertainties discussed in Sec.~\ref{sec:systematics}.
To account for the impact of signal contamination, we correct the LM4 and LM8 yields by subtracting the expected
increase in the OF background estimate that would occur if these signals were present in the data.
As mentioned above, the contribution from LM4 and LM8 to the \MET\ template background estimate is negligible.
The expected LM4 and LM8 yields exceed the upper limits on the non-SM contributions to the yields in those signal regions
with a minimum \MET\ requirement of 200\GeV.

\begin{table*}[tbhp]
\begin{center}
\footnotesize
\topcaption{\label{resultsyieldtable}
Summary of results in the regions \MET $>$ 30, 60, 100, 200, and 300\GeV for \njets\ $\ge$~2.
The total predicted background (total bkg) is the sum of the
\zjets\ background predicted from the \gjets\ \MET template method (\Z bkg), the background predicted from OF
events (OF bkg), and the $\PW\Z+\Z\Z$ background predicted from simulation (VZ bkg).
The first (second) uncertainty indicates the statistical (systematic) component.
For the observed yield (data), the first (second) number in parentheses is the yield in the $\Pe\Pe$ ($\Pgm\Pgm$) final state.
The 95\% CL observed and expected upper limits (UL) on the non-SM yield are indicated.
The expected NLO yields for the LM4 and LM8 benchmark SUSY scenarios are also given,
including the systematic uncertainties and the correction for the impact of signal contamination indicated in the text.
}
\begin{tabular}{l|ccccc}

\hline
\hline
              &   \MET $>30$\GeV   & \MET $>60$\GeV   & \MET $>100$\GeV   &   \MET $>200$\GeV  &  \MET $>$ 300\GeV \\
\hline		

\Z bkg       &  15070 $\pm$  161 $\pm$ 4822  &   484 $\pm$   23 $\pm$  155  &    36 $\pm$    4.6 $\pm$   11  &     2.4 $\pm$    0.6 $\pm$    0.8  &     0.4 $\pm$  0.2 $\pm$  0.1 \\
OF bkg       &   1116 $\pm$   13 $\pm$  100  &   680 $\pm$   10 $\pm$   61  &   227 $\pm$    6.0 $\pm$   20  &      11 $\pm$    1.3 $\pm$    3.1  &     1.6 $\pm$  0.5 $\pm$  0.4 \\
VZ bkg       &   269 $\pm$   0.9 $\pm$  135  &    84 $\pm$  1.0 $\pm$   42  &    35 $\pm$    0.5 $\pm$   17  &     5.3 $\pm$    0.4 $\pm$    2.7  &     1.2 $\pm$  0.4 $\pm$  0.6 \\
\hline
\hline
Total bkg   & 16455 $\pm$   161 $\pm$ 4825  &  1249 $\pm$   25 $\pm$  172  &   297 $\pm$    7.5 $\pm$   29  &    19   $\pm$    1.5 $\pm$    4.1  &     3.2 $\pm$  0.7 $\pm$  0.7 \\
Data          &    16483 (8243,8240)&    1169 (615,554)  &    290 (142,148)   &     14 (8,6)          &    0 \\
\hline
\hline
Observed UL           & 9504           &  300           &   57            &     8.3          &   3.0         \\
Expected UL           & 9478           &  349           &   60            &      11          &   4.6         \\

LM4                   & $120 \pm 7.0$  &  $108 \pm 6.7$  &  $93 \pm 6.6$  &  $53 \pm 7.3$  &  $24 \pm 6.2$   \\
LM8                   &  $52 \pm 3.2$  &   $46 \pm 3.0$  &  $37 \pm 2.8$  &  $21 \pm 2.8$  &  $9.1 \pm 2.3$  \\

\hline
\hline
\end{tabular}
\end{center}
\end{table*}

\begin{table*}[bth]
\begin{center}
\footnotesize
\topcaption{\label{resultsyieldtable3}
Summary of results for \njets\ $\ge$~3. The details are the same as for the \njets\ $\ge$~2
results quoted in Table~\ref{resultsyieldtable}, except that the total background prediction is based on the
average of the background predictions from the QCD and \gjets\ template methods, which are quoted separately.
}
\begin{tabular}{l|ccccc}

\hline
\hline
                       &   \MET $>30$\GeV  & \MET $>60$\GeV   & \MET $>100$\GeV   &   \MET $>200$\GeV  &  \MET $>$ 300\GeV   \\

\hline
\Z bkg (QCD)   &  4010 $\pm$ 65 $\pm$ 800      & 191 $\pm$  12 $\pm$ 56       &  11 $\pm$ 0.7 $\pm$  11       & 0.7 $\pm$ 0.05 $\pm$ 0.7 & 0.1 $\pm$ 0.02 $\pm$  0.1 \\
\Z bkg (\gjets)& 3906 $\pm$  61 $\pm$ 1250     & 187 $\pm$  10 $\pm$ 60       &  14 $\pm$ 1.7 $\pm$ 4.6       & 1.7 $\pm$  0.5 $\pm$ 0.5 & 0.3 $\pm$  0.2 $\pm$  0.1 \\
OF bkg         &  442 $\pm$ 8.0 $\pm$   40     & 284 $\pm$ 7.0 $\pm$ 26       & 107 $\pm$ 4.1 $\pm$  10       & 7.5 $\pm$  1.1 $\pm$ 2.0 & 1.1 $\pm$  0.4 $\pm$  0.3 \\
WZ bkg         &   86 $\pm$ 1.0 $\pm$   43     &  26 $\pm$ 0.3 $\pm$ 13       &  11 $\pm$ 0.2 $\pm$ 5.6       & 1.9 $\pm$  0.2 $\pm$ 1.0 & 0.4 $\pm$  0.2 $\pm$  0.2 \\

\hline
Total bkg (QCD)        &  4539 $\pm$ 66 $\pm$ 802  &   502 $\pm$   14 $\pm$   63  &   129 $\pm$ 4.2 $\pm$   16    & 10 $\pm$ 1.1 $\pm$ 2.3  & 1.6 $\pm$  0.4 $\pm$  0.4 \\
Total bkg (\gjets)     &  4435 $\pm$ 62 $\pm$ 1251 &   498 $\pm$   12 $\pm$   66  &   132 $\pm$ 4.4 $\pm$   12    & 11 $\pm$ 1.2 $\pm$ 2.2  & 1.9 $\pm$  0.5 $\pm$  0.4 \\
\hline
\hline
    Total bkg (average)& 4487 $\pm$ 64 $\pm$ 1027  &   500 $\pm$ 13 $\pm$ 65      & 131 $\pm$ 4.3 $\pm$ 14        & 11 $\pm$ 1.2 $\pm$ 2.3  & 1.8 $\pm$ 0.5 $\pm$ 0.4  \\
    Data               &    4501 (2272,2229)       &               479 (267,212)  &    137 (73,64)                &        8 (3,5)       &          0        \\
\hline
\hline

Observed UL                   &     2028          &         120       &       40         &         6.7        &     3.0            \\
Expected UL                   &     2017          &         134       &       36         &         8.4        &     3.9            \\

LM4                           &     $97 \pm 6.1$  &     $90 \pm 6.1$  &    $79 \pm 6.6$  &      $44 \pm 7.1$  &   $19 \pm 5.4$     \\
LM8                           &     $42 \pm 2.6$  &     $39 \pm 2.5$  &    $33 \pm 2.5$  &      $19 \pm 2.7$  &  $8.3 \pm 2.1$     \\

\hline
\hline

\end{tabular}
\end{center}
\end{table*}

\section{Signal Acceptance and Efficiency Uncertainties}
\label{sec:systematics}

The acceptance and efficiency, as well as the systematic uncertainties on these quantities,
depend on the signal model under consideration.
For some of the individual uncertainties,
we quote values based on SM control samples with kinematic properties similar to the SUSY benchmark models.
For others that depend strongly on the kinematic properties of the event, the systematic
uncertainties are quoted model-by-model and separately for the various signal regions.

The systematic uncertainty on the lepton acceptance consists
of two parts: the trigger efficiency uncertainty and the
identification and isolation uncertainty. The trigger efficiency
for two leptons of $\pt>20\GeV$ is measured in a $\Z \to \ell\ell$ data sample,
with an uncertainty of 2\%. We verify that the simulation reproduces the lepton identification
and isolation efficiencies in data using
$\Z \to \ell\ell$ samples, within a systematic uncertainty of 2\% per lepton.

Another significant source of systematic uncertainty in the acceptance is
associated with the jet and \MET energy scale.  The impact
of this uncertainty depends on the final state under consideration.
Final states characterized by very large \MET are less sensitive to this uncertainty
than those with \MET values near the minimum signal region requirements.
To estimate this uncertainty, we have used the method of Ref.~\cite{ref:top}
to evaluate the systematic uncertainties in the acceptance for the two benchmark SUSY points.
The energies of jets in this analysis are known to 7.5\% (not all the corrections in Ref.~\cite{ref:jes} were applied).
For LM4 and LM8, the corresponding systematic uncertainties on the signal region yields vary from 4--6\% for \MET $>$ 100\GeV
to 24--28\% for \MET $>$ 300\GeV.

The impact of the hadronic scale uncertainty on the \JZB efficiency is estimated by
varying the jet energy scale by one standard deviation~\cite{ref:jes}. This  leads
to a systematic uncertainty of 3--6\% on the signal efficiency, depending on the model and the signal region.
The \JZB scale is then varied by 5\% to account for the uncertainty in unclustered energy deposits.
The corresponding signal efficiency uncertainties vary between 1\% ($\JZB>50\GeV$) and 7\% ($\JZB>250\GeV$) for LM4,
and between 1\% and 10\% for LM8.

Uncertainties on the PDFs are determined individually for each
scenario and are propagated to the efficiency, as recommended in Ref.~\cite{Bourilkov:2006cj}.
The uncertainty associated with the integrated luminosity is 2.2\%~\cite{ref:lumi}.

\section{Interpretation}\label{sec:interpret}

In the absence of a significant excess, we set upper limits
on the production cross section of SMS models~\cite{Alves:2011wf,ArkaniHamed:2007fw,SMS1},
which represent decay chains of new particles that may occur in a wide variety of BSM physics scenarios, including SUSY.
We provide the signal selection efficiencies in the model parameter space.
These efficiencies may be employed to validate and calibrate the results of fast
simulation software used to determine the signal efficiency of an arbitrary BSM model.
This allows our results to be applied to BSM models beyond those examined in this paper.
We also provide cross section upper limits in the parameter space of these models, and exclude a region of the parameter space
assuming reference cross sections and a 100\% branching fraction to the final state under consideration (the \Z boson
is allowed to decay according to the well-known SM branching fractions).

\begin{figure}[hbtp]
  \begin{center}
    \includegraphics[width=0.4\textwidth]{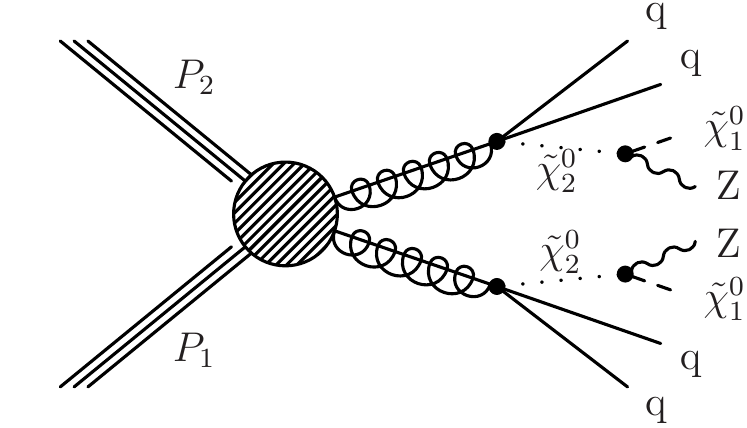}
    \caption{Simplified model for the production of two gluinos decaying into two \Z bosons, two $\chiz_1$ particles, and jets.}
    \label{fig:T5zz}
  \end{center}
\end{figure}

Figure~\ref{fig:T5zz} illustrates the process considered in this study: two gluinos are produced,
each of which decays to a pair of jets and the second-lightest neutralino $\chiz_{2}$, which itself decays
to a \Z boson and the LSP $\chiz_{1}$.
The parameters of the model are the masses of the gluino ($m_{\sGlu}$) and of the LSP ($m_{\chiz_{1}}$).
The mass of the intermediate neutralino ($m_{\chiz_{2}}$) is fixed to
$m_{\chiz_{2}} = m_{\chiz_{1}} + x\cdot(m_{\sGlu}-m_{\chiz_{1}})$, with $x=0.5$.
The results are only presented in the region where the particle masses as specified above satisfy $m_{\chiz_{2}}>m_{\chiz_{1}}+m_{\Z}$.
Additional interpretations for a different choice of $x$ as well as for
a model inspired by gauge-mediated SUSY breaking are included in the supplementary materials of this paper.

For the JZB analysis, we calculate the observed and expected upper limits on the cross section
using the results in all signal regions, and select the observed limit corresponding to the best expected limit
for each parameter point.
For the MET analysis, the cross section upper limit is based on simultaneous counting experiments in the three exclusive regions of
100\GeV $<$ \MET $<$ 200\GeV, 200\GeV $<$ \MET\ $<$ 300\GeV, and \MET $>$ 300\GeV,
as summarized in Table~\ref{resultsyieldtable2}, since this exclusive binning improves the sensitivity
to a specific BSM model.
The model-dependent systematic uncertainties (energy scale and PDF uncertainties) are determined for each point.
To interpret these limits in terms of the gluino pair-production cross section, we use a reference cross section $\sigma^{\text{NLO-QCD}}$
and determine the 95\% CL exclusion contours at 1/3, 1, and 3 times $\sigma^{\text{NLO-QCD}}$, to establish
how the limit changes with the cross section. This reference cross section $\sigma^{\text{NLO-QCD}}$
corresponds to gluino pair-production in the limit of infinitely heavy squarks, calculated at NLO
using \PROSPINO~\cite{Beenakker-1996} and the CTEQ6~\cite{Pumplin:2002vw} PDFs.

\begin{table*}[thbp]
\begin{center}
\footnotesize
\topcaption{\label{resultsyieldtable2}
Summary of results for the
\MET\ template analysis in the exclusive regions
100\GeV $<$ \MET $<$ 200\GeV, 200\GeV $<$ \MET\ $<$ 300\GeV, and \MET $>$ 300\GeV
for \njets\ $\ge$ 2 used for the SMS
exclusions of Section~\ref{sec:interpret}.
The total predicted background (total bkg) is the sum of the
\zjets\ background predicted from the \gjets\ \MET templates method (\Z bkg), the  background predicted from opposite-flavor
events (OF bkg), and the $\PW\Z+\Z\Z$ background predicted from simulation (VZ bkg).
The uncertainties include both the statistical and systematic contributions.
For the observed yield (data), the first (second) number in parentheses is the yield in the $\Pe\Pe$ ($\Pgm\Pgm$) final state.
}
\begin{tabular}{l|ccc}
\hline
\hline
               & 100\GeV $<$ \MET $<$ 200\GeV & 200\GeV $<$ \MET $<$ 300\GeV &  \MET $>$ 300\GeV \\

\hline
\Z bkg       &     33 $\pm$    4.5 $\pm$   11  &   1.9 $\pm$    0.5 $\pm$    0.6 &   0.4 $\pm$  0.2 $\pm$  0.1 \\
OF bkg       &    215 $\pm$    5.8 $\pm$   19  &   10  $\pm$    1.2 $\pm$    2.7 &   1.6 $\pm$  0.5 $\pm$  0.4 \\
VZ bkg       &     29 $\pm$    0.2 $\pm$   15  &   4.2 $\pm$    0.1 $\pm$    2.1 &   1.2 $\pm$  0.4 $\pm$  0.6 \\
\hline
\hline
Total bkg    &    278 $\pm$    7.4 $\pm$   27  &    16 $\pm$    1.3 $\pm$   3.5  &   3.2 $\pm$  0.7 $\pm$  0.7 \\
Data         &        276 (134,142)            &      14 (8,6)                   &                           0 \\
\hline
\hline

\end{tabular}
\end{center}
\end{table*}

\begin{figure}[htbp]
  \begin{center}
    \includegraphics[width=0.49\textwidth]{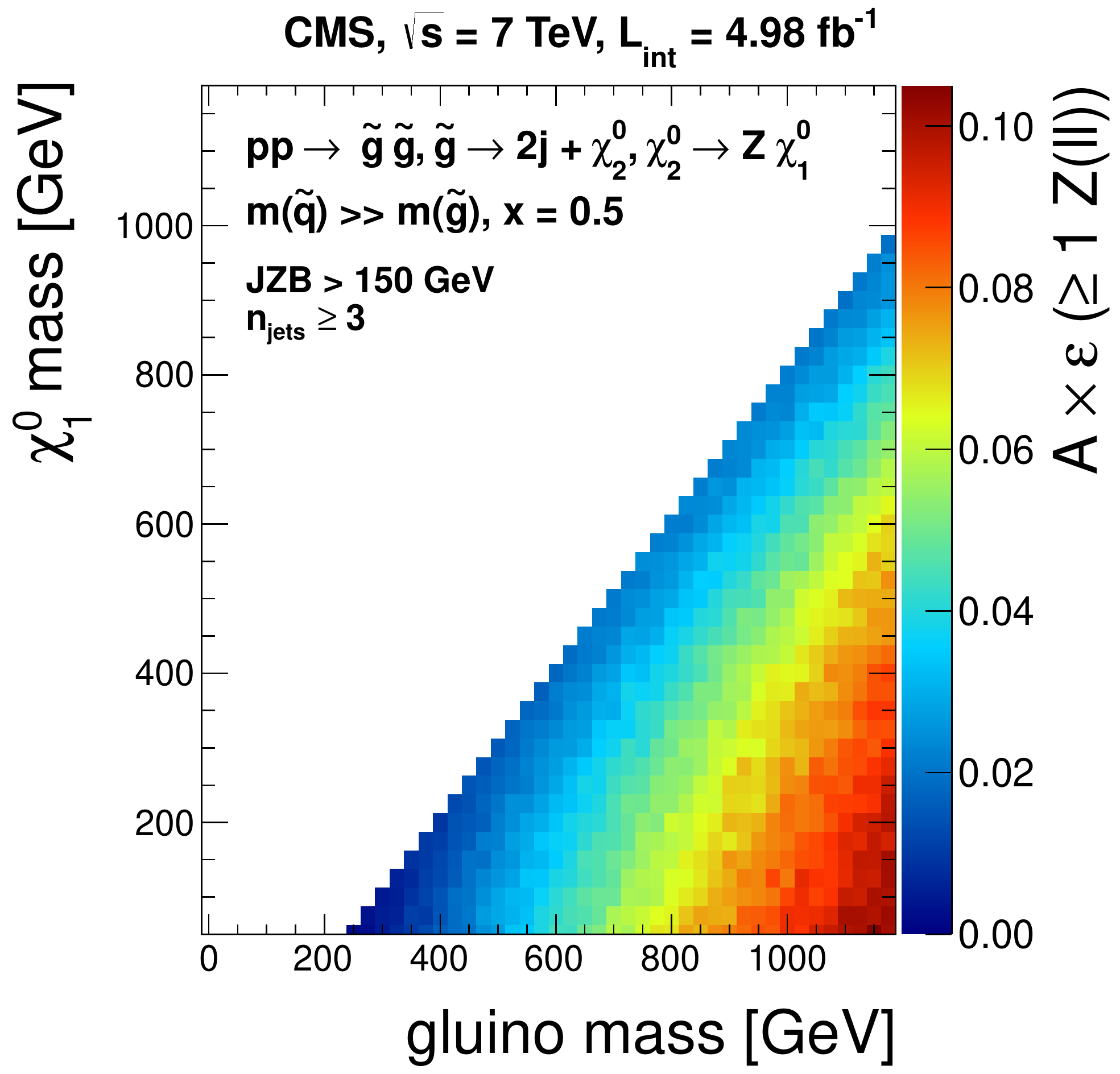}
    \includegraphics[width=0.49\textwidth]{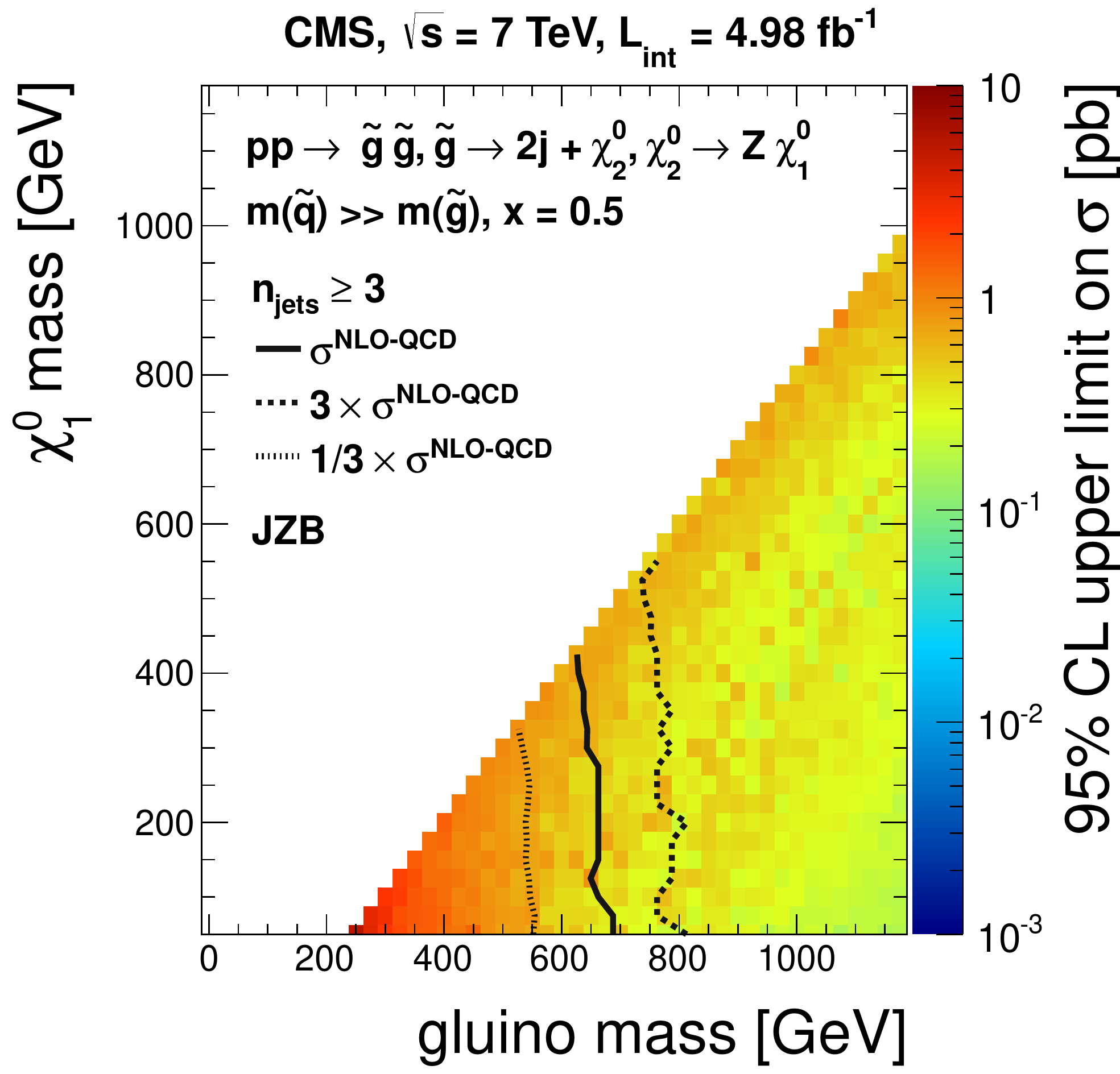}
    \caption{\footnotesize
Limits on the SMS topology described in the text, based on the \JZB method: (\cmsLeft) signal efficiency times acceptance normalized
to the number of events with at least one $\Z\to\ell\ell$ decay for the $\JZB>150\GeV$ region;
(\cmsRight) 95\% CL upper limits on the  total gluino pair-production cross section.
The region to the left of the solid contour is excluded assuming that the gluino pair-production
cross section is $\sigma^{\text{NLO-QCD}}$, and that the branching fraction to this SMS topology is 100\%.
The dotted and dashed contours indicate the excluded region when the cross section is varied by a factor of three.
The signal contribution to the control regions is taken into account.
}
    \label{fig:SMS05JZB}
  \end{center}
\end{figure}

Figure~\ref{fig:SMS05JZB} shows the signal selection efficiency times acceptance for the \JZB\ $>$ 150\GeV signal region
for the topology described above, normalized
to the number of events with at least one leptonically-decaying \Z.
The 95\% CL upper limits on the total gluino pair-production cross section are also shown.
The $\JZB>250\GeV$ region has the best sensitivity throughout most of the parameter space of this model.
The signal contribution to the \zjets control sample has been taken into account in these limits.
In this mass spectrum, the \Z boson and \MET directions are weakly correlated and the sensitivity
of the \JZB search is reduced at low LSP masses.

\begin{figure}[hbtp]
  \begin{center}

    \includegraphics[width=0.49\textwidth]{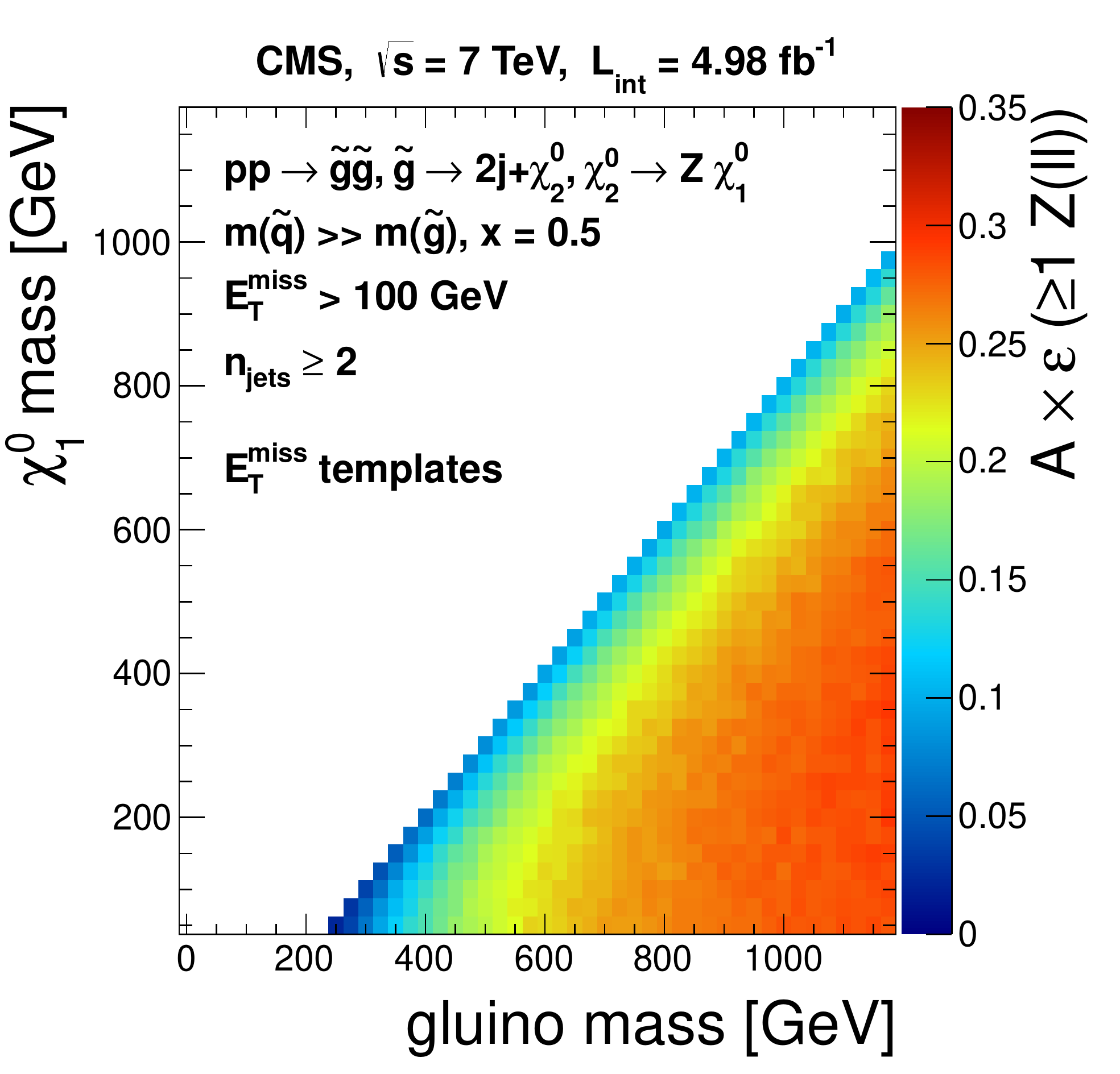}
    \includegraphics[width=0.49\textwidth]{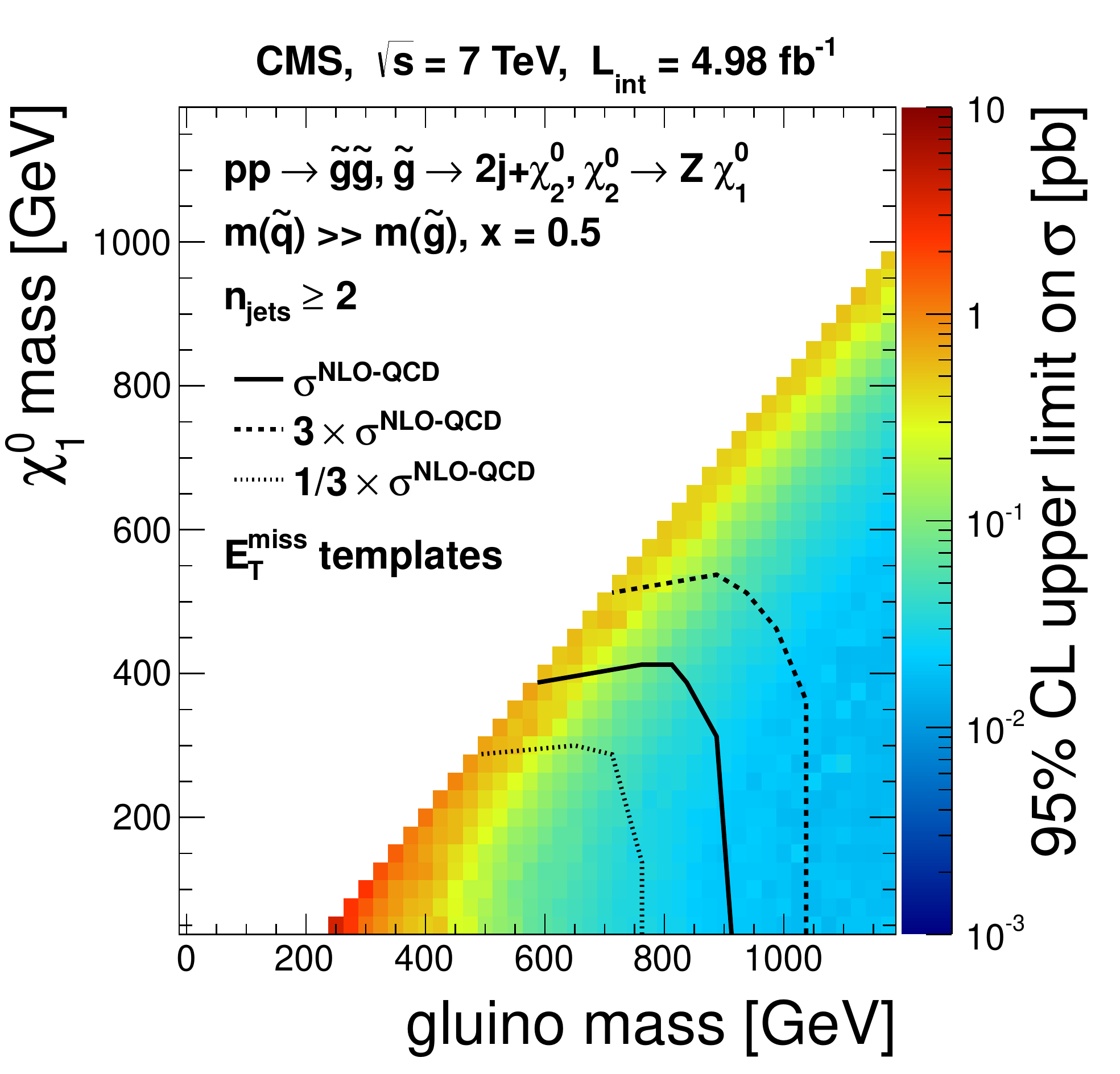}
    \caption{\footnotesize
Limits on the SMS topology described in the text, based on the \MET\ template
method: (\cmsLeft) signal efficiency times acceptance normalized
to the number of events with at least one $\Z\to\ell\ell$ decay for the $\MET>100\GeV$ region;
(\cmsRight) 95\% CL upper limits on the total gluino pair-production cross section.
The region to the left of the solid contour is excluded assuming that the gluino pair-production
cross section is $\sigma^{\text{NLO-QCD}}$, and that the branching fraction to this SMS topology is 100\%.
The dotted and dashed contours indicate the excluded region when the cross section is varied by a factor of three.
The signal contribution to the control regions is negligible.
}
    \label{fig:SMStemplates}
  \end{center}
\end{figure}

Figure~\ref{fig:SMStemplates} shows the signal selection efficiency times acceptance for the $\MET > 100$\GeV signal region in the \MET\ template analysis,
normalized to the number of events with at least one leptonically-decaying \Z.
The 95\% CL upper limits on the total gluino pair-production cross section,
based on the three simultaneous counting experiments in the regions
100\GeV $<$ \MET $<$ 200\GeV, 200 $<$ \MET $<$ 300\GeV, and \MET $>$ 300\GeV,
are also shown. The signal contribution to the QCD and \gjets\ control samples used to estimate the \Z background and to the $\Pe\Pgm$
control sample used to estimate the flavor-symmetric background is negligible.
This interpretation is based on the results with \njets\ $\ge$ 2; we find comparable
results using \njets\ $\ge$ 3.

\section{Additional Information for Model Testing}
\label{sec:outreach}

\begin{figure}[hbtp]
  \begin{center}
    \includegraphics[width=0.49\textwidth]{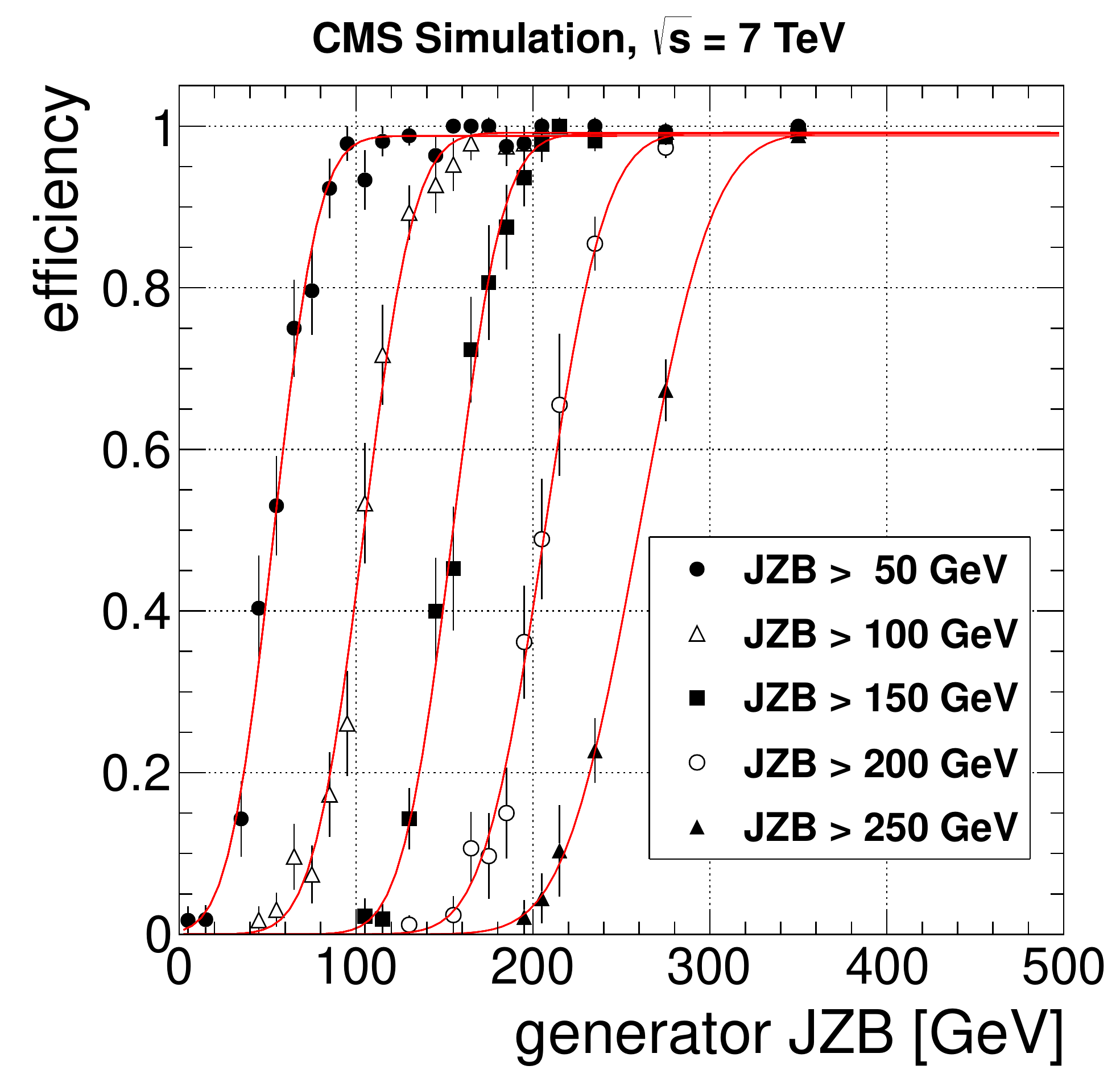}
    \includegraphics[width=0.49\textwidth]{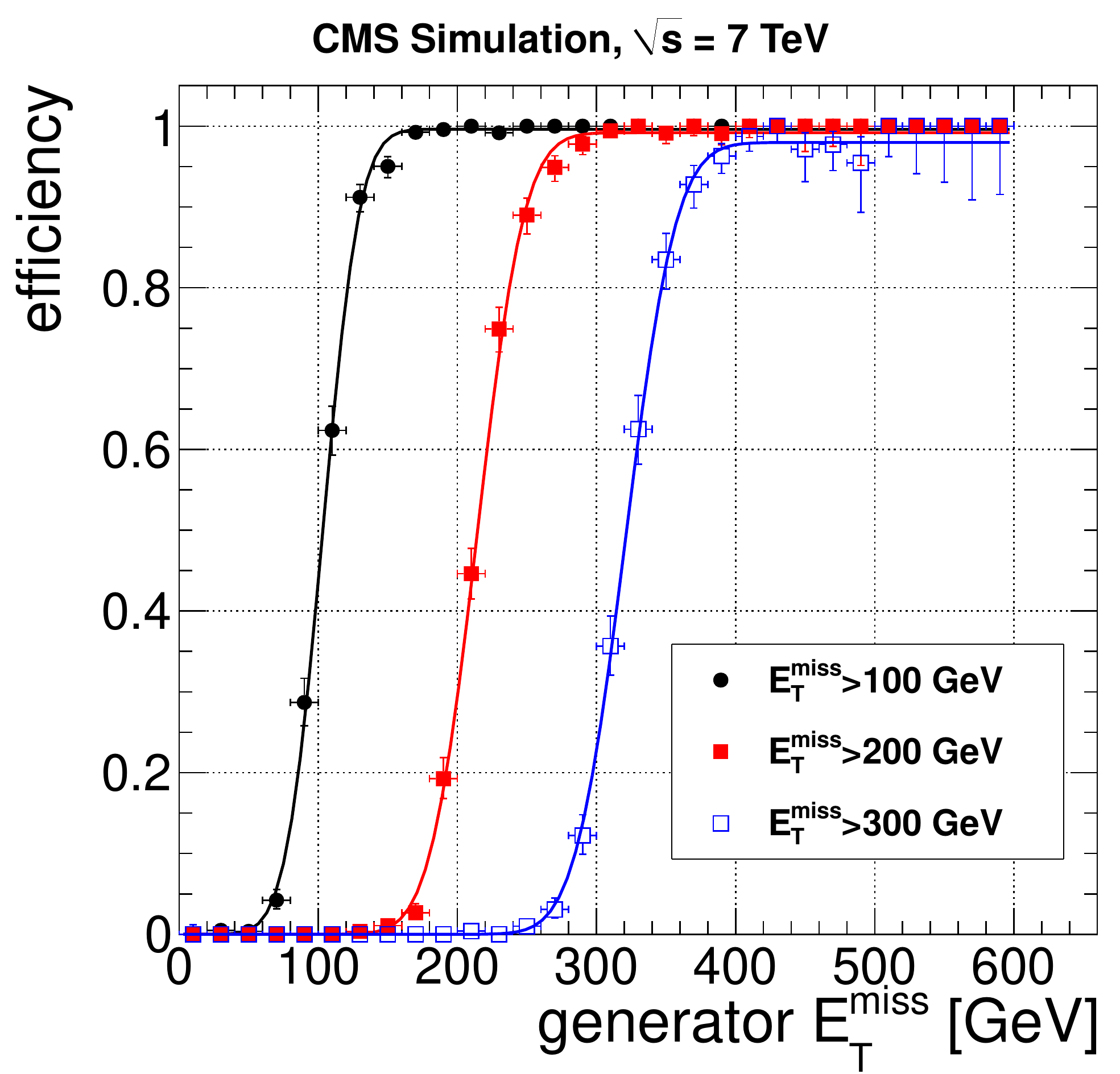}
    \caption{Reconstructed \JZB (\cmsLeft) and \MET\ (\cmsRight) selection efficiencies as a function of the generator-level quantity,
    for the different signal regions in the LM4 simulation.}
    \label{fig:JZBresp}
  \end{center}
\end{figure}

Other models of BSM physics in the dilepton final state can be constrained in an approximate manner by simple
generator-level studies that compare the expected number of events in \lumifinal\
with the upper limits from Sections~\ref{ssec:jzbresults} and~\ref{sec:met_results}.
The key ingredients of such studies are the kinematic requirements described
in this paper, the lepton efficiencies, and the detector responses for \MET and \JZB.
The  trigger efficiencies for events containing $\Pe\Pe$, $\Pe\Pgm$, or $\Pgm\Pgm$ lepton pairs
are 100\%, 95\%, and 90\%, respectively.
The muon identification efficiency is approximately $91\%$;
the electron identification efficiency varies approximately linearly from about 83\% at
$\pt = 20\GeV$ to about 93\% for $\pt > 60\GeV$ and then is flat.
The lepton isolation efficiency depends on the lepton momentum, as well as on the jet activity in the event.
In \ttbar events, the efficiency varies approximately linearly from about 85\% (muons)
and 88\% (electrons) at $\pt=20\GeV$ to about 97\% for $\pt>60\GeV$.
In LM4 (LM8) events, this efficiency is decreased by approximately 5\% (10\%) over the whole momentum spectrum.
The average detector response for \JZB is $92\%$. In order to better quantify the \JZB and \MET\ selection efficiencies,
we study the probability for an event to pass a given reconstructed \JZB or \MET\ requirement as a function of the
generator-level quantity. Here, generator-level \MET\ is the negative vector sum of the stable, invisible particles,
including neutrinos and SUSY LSP's.
The response is parametrized by a function of the form (see Fig.~\ref{fig:JZBresp}):
\begin{equation}
\varepsilon(x)= \varepsilon_\text{plateau}\frac{1}{2}
\left[\mathrm{erf}\left(\frac{x-x_{\mathrm{thresh}}}{\sigma}\right)+1\right].
\end{equation}
The fitted parameters are summarised in Table~\ref{tab:JZBresp}.

To approximate the requirement on the jet multiplicity, we count quarks or gluons from the hard scattering process
that satisfy the acceptance requirements $\pt > 30\GeV$ and $|\eta|<3.0$. We have tested this efficiency model
with the LM4 and LM8 benchmark models, and find that the efficiency from our model is consistent with the
expectation from the full reconstruction to within about 15\%.

\begin{table}[hbtp]
\renewcommand{\arraystretch}{1.3}
\begin{center}
\topcaption{Parameters of the \JZB (top) and \MET\ (bottom) response function. The parameter $\sigma$ is the resolution,
$x_\text{thresh}$ is the \JZB or \MET\ value at the center of the efficiency curve,
and $\varepsilon_\text{plateau}$ is the efficiency on the plateau.}\label{tab:JZBresp}
\begin{tabular}{  l   c  c c }
\hline
\hline
Region          & $\sigma$ [\GeV]    &$x_\text{thresh}$ [\GeV] & $\varepsilon_\text{plateau}$\\
\hline
$\JZB>50\GeV$   & $30$     & $ 55$ & $0.99$ \\
$\JZB>100\GeV$  & $30$     & $108$ & $0.99$ \\
$\JZB>150\GeV$  & $32$     & $156$ & $0.99$ \\
$\JZB>200\GeV$  & $39$     & $209$ & $0.99$ \\
$\JZB>250\GeV$  & $45$     & $261$ & $0.98$ \\
\hline
\hline
$\MET>100\GeV$  &  29                &    103          &     1.00          \\
$\MET>200\GeV$  &  38                &    214          &     0.99          \\
$\MET>300\GeV$  &  40                &    321          &     0.98          \\
\hline
\hline
\end{tabular}
\end{center}
\end{table}

\section{Summary}\label{sec:conclusion}

We have performed a search for BSM physics in final states with a leptonically-decaying \Z boson,
jets, and missing transverse energy.
Two complementary  strategies are used to suppress the dominant
\zjets background and to estimate the remaining background from data control samples:
the jet-$\Z$ balance method and the \MET\ template method. Backgrounds from \ttbar processes
are estimated using opposite-flavor lepton pairs and dilepton invariant mass sidebands.
We find no evidence for anomalous yields beyond standard model (SM) expectations and place upper limits
on the non-SM contributions to the yields in the signal regions.
The results are interpreted in the context of simplified model spectra.
We also provide information on the detector response and efficiencies to
allow tests of BSM models with \Z bosons that are not considered in the present study.

\section*{Acknowledgments}

We wish to congratulate our colleagues in the CERN accelerator departments for the excellent performance of the LHC machine. We thank the technical and administrative staff at CERN and other CMS institutes, and acknowledge support from: FMSR (Austria); FNRS and FWO (Belgium); CNPq, CAPES, FAPERJ, and FAPESP (Brazil); MES (Bulgaria); CERN; CAS, MoST, and NSFC (China); COLCIENCIAS (Colombia); MSES (Croatia); RPF (Cyprus); MoER, SF0690030s09 and ERDF (Estonia); Academy of Finland, MEC, and HIP (Finland); CEA and CNRS/IN2P3 (France); BMBF, DFG, and HGF (Germany); GSRT (Greece); OTKA and NKTH (Hungary); DAE and DST (India); IPM (Iran); SFI (Ireland); INFN (Italy); NRF and WCU (Korea); LAS (Lithuania); CINVESTAV, CONACYT, SEP, and UASLP-FAI (Mexico); MSI (New Zealand); PAEC (Pakistan); MSHE and NSC (Poland); FCT (Portugal); JINR (Armenia, Belarus, Georgia, Ukraine, Uzbekistan); MON, RosAtom, RAS and RFBR (Russia); MSTD (Serbia); MICINN and CPAN (Spain); Swiss Funding Agencies (Switzerland); NSC (Taipei); TUBITAK and TAEK (Turkey); STFC (United Kingdom); DOE and NSF (USA). Individuals have received support from the Marie-Curie programme and the European Research Council (European Union); the Leventis Foundation; the A. P. Sloan Foundation; the Alexander von Humboldt Foundation; the Belgian Federal Science Policy Office; the Fonds pour la Formation \`a la Recherche dans l'Industrie et dans l'Agriculture (FRIA-Belgium); the Agentschap voor Innovatie door Wetenschap en Technologie (IWT-Belgium); the Council of Science and Industrial Research, India; and the HOMING PLUS programme of Foundation for Polish Science, cofinanced from European Union, Regional Development Fund.

\ifthenelse{\boolean{cms@external}}{\section*{References}}{}
\bibliography{auto_generated}   
\ifthenelse{\boolean{cms@external}}{}{
\appendix
\clearpage

\section{Additional Interpretation of the Results}\label{sec:smsapp}

In this appendix we interpret our results in the context of two additional SMS topologies. The first topology is the
same as discussed in Sec.~7, in which the LSP is the lightest neutralino, but with a different choice of the $\chiz_2$ mass parameter, $x=0.75$,
so that the $\chiz_{2}$ is closer in mass to the gluino than to the LSP.
The second is a topology inspired by gauge-mediated SUSY-breaking (GMSB) models, in which the LSP is a light gravitino (mass $\lesssim$ 1 keV),
which is treated here as massless.
In this scenario, we consider gluino pair-production where each gluino decays to a pair of jets and the lightest neutralino $\chiz_1$, which
itself decays to a \Z boson and the gravitino ($\tilde{G}$) LSP, as shown in Fig.~\ref{fig:gmsb}.
If the $\chiz_{1}$ is mostly bino then the decay $\chiz_{1}\rightarrow\gamma~\tilde{G}$ dominates,
while the decay $\chiz_{1} \rightarrow \Z~\tilde{G}$ can become favored if the $\chiz_1$ is mostly wino or higgsino.
The parameters of this model are the masses of the gluino and of the lightest neutralino $\chiz_1$.

Results for the neutralino LSP scenario are presented in Fig.~\ref{fig:SMS05JZB3} (JZB analysis) and Fig.~\ref{fig:T5zzl} (MET analysis).
Results for the gravitino LSP scenario are presented in Fig.~\ref{fig:SMS05JZB2} (JZB analysis) and Fig.~\ref{fig:T5zzgmsb} (MET analysis).

The \JZB search relies on the correlation between the \Z boson and the \MET directions, which leads to an asymmetry in the \JZB distribution.
The sensitivity of this search is thus reduced in mass spectra that lead to symmetric \JZB, as can be the case in the GMSB-inspired
scenario in the region of parameter space that is evident, e.g., in Fig.~\ref{fig:SMS05JZB2}.

\begin{figure}[hbtp]
  \begin{center}
    \includegraphics[width=0.4\textwidth]{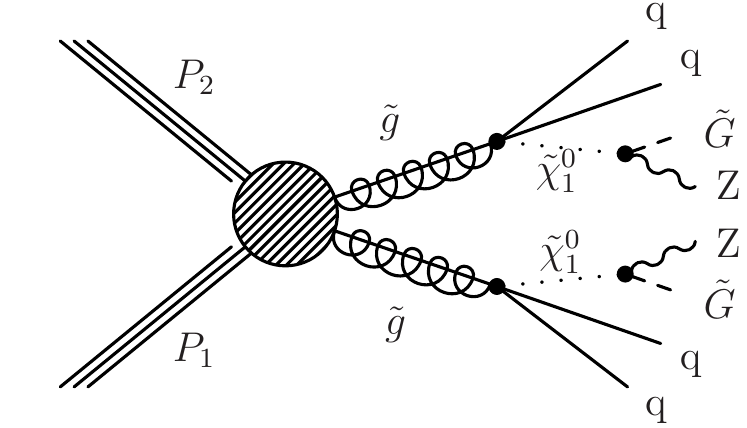}
    \caption{Simplified model for the production of two gluinos decaying into two \Z bosons, two gravitinos, and jets.
    \label{fig:gmsb}}
  \end{center}
\end{figure}

\clearpage

\begin{figure}[!htbp]
  \begin{center}
    \includegraphics[width=0.48\textwidth]{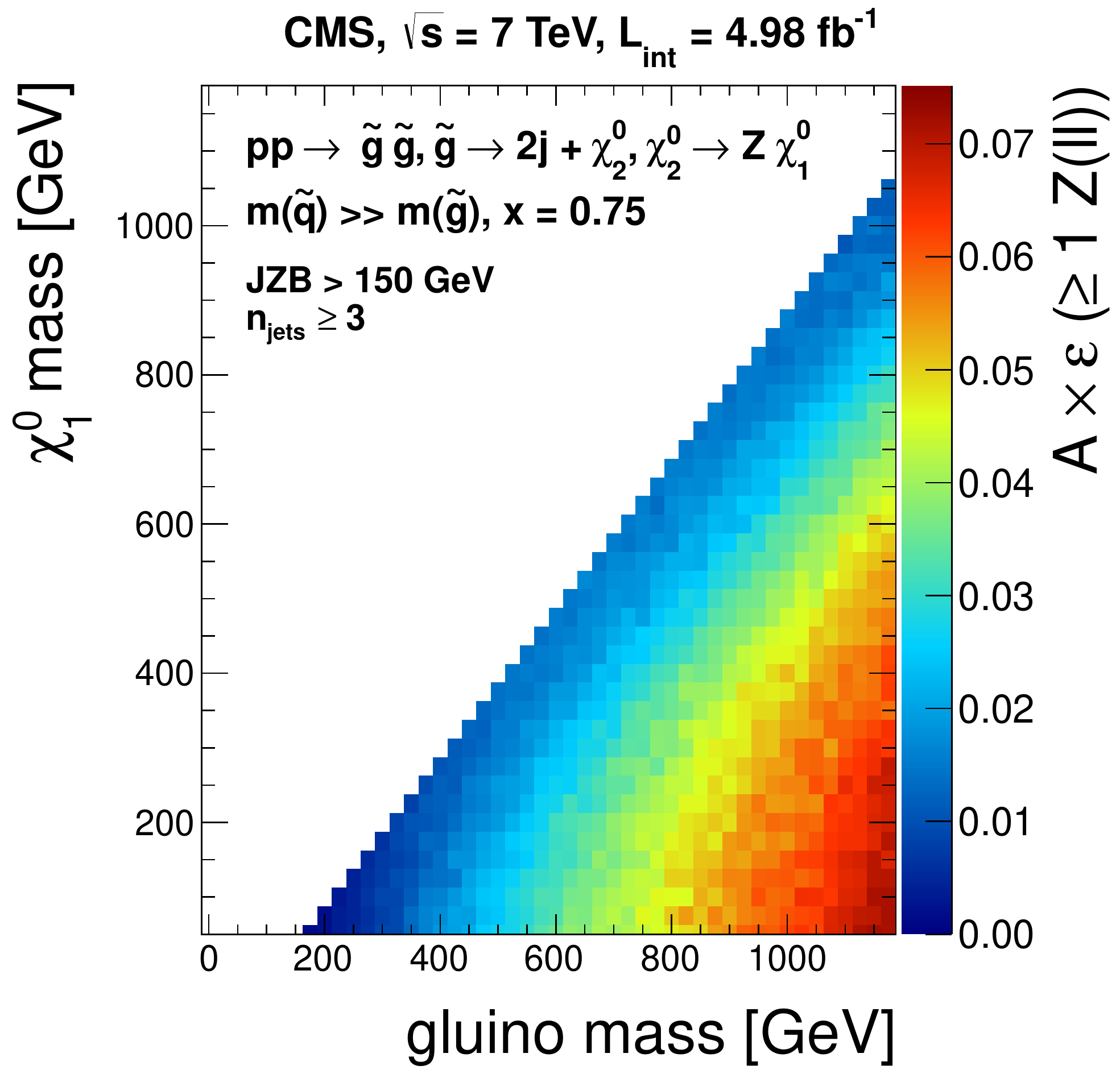}
    \includegraphics[width=0.48\textwidth]{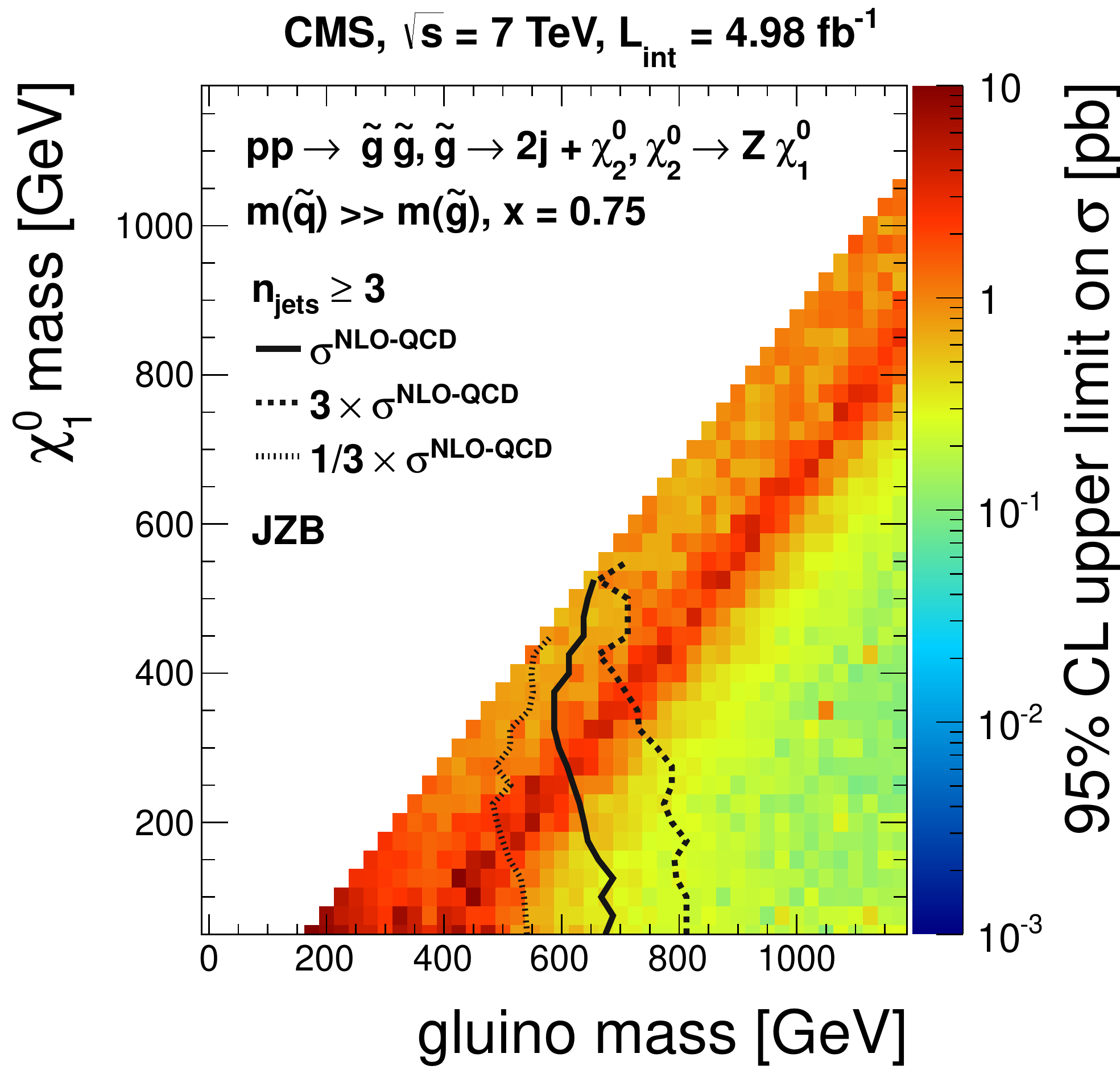}
    \caption{
\footnotesize
Limits on the SMS topology with neutralino LSP ($x=0.75$), based on the \JZB method: (left)
signal efficiency times acceptance normalized
to the number of events with at least one $Z\to\ell\ell$ decay for the $\JZB>150\GeV$ region;
(right) 95\% CL upper limits on the  total gluino pair-production cross section.
The region to the left of the solid contour is excluded assuming that the gluino pair-production
cross section is $\sigma^{\text{NLO-QCD}}$, and that the branching fraction to this SMS topology is 100\%.
The dotted and dashed contours indicate the excluded region when the cross section is varied by a factor of three.
The signal contribution to the control regions is taken into account.
}
    \label{fig:SMS05JZB3}
  \end{center}
\end{figure}

\begin{figure}[!htbp]
  \begin{center}
    \includegraphics[width=0.48\textwidth]{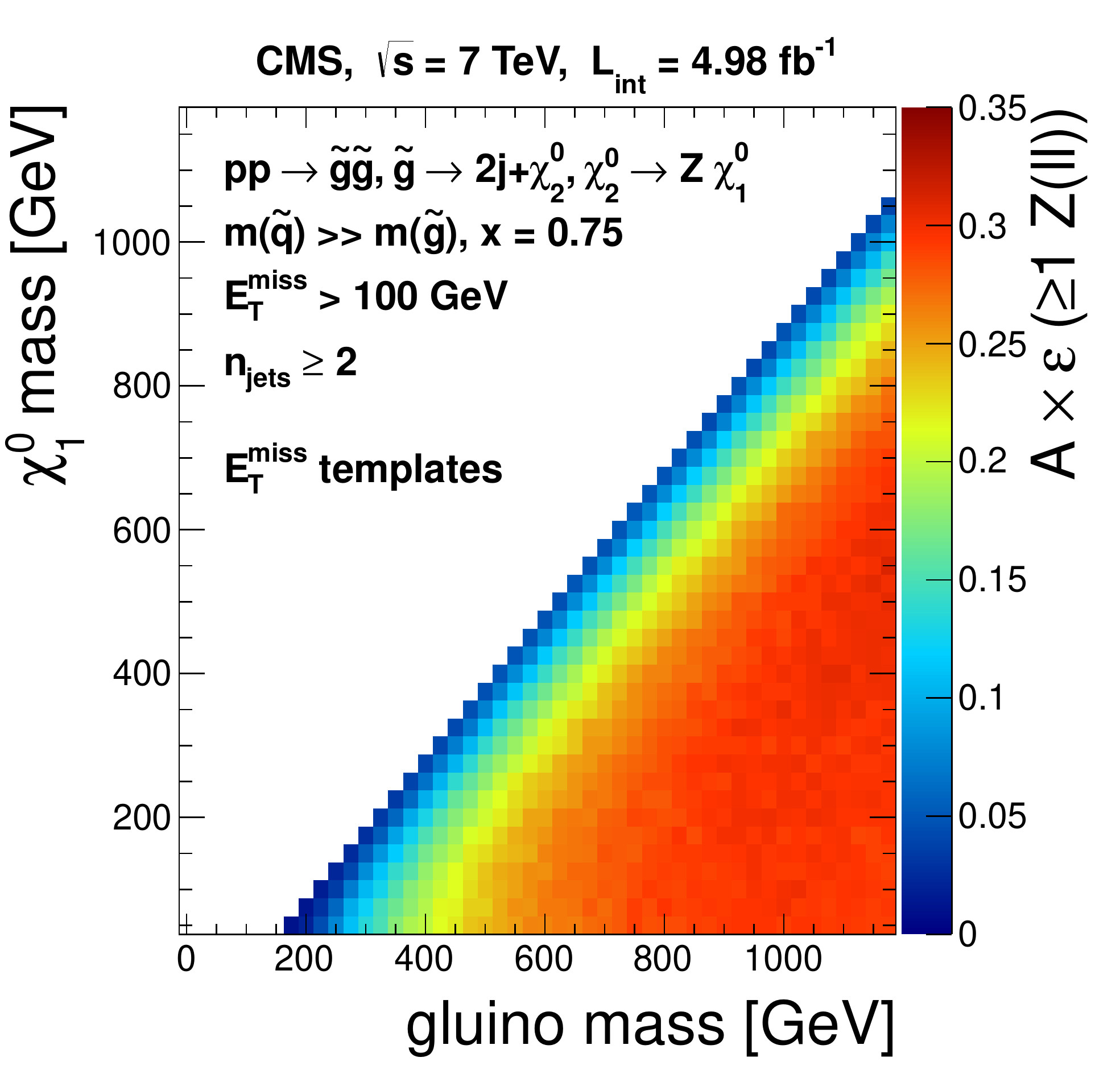}
    \includegraphics[width=0.48\textwidth]{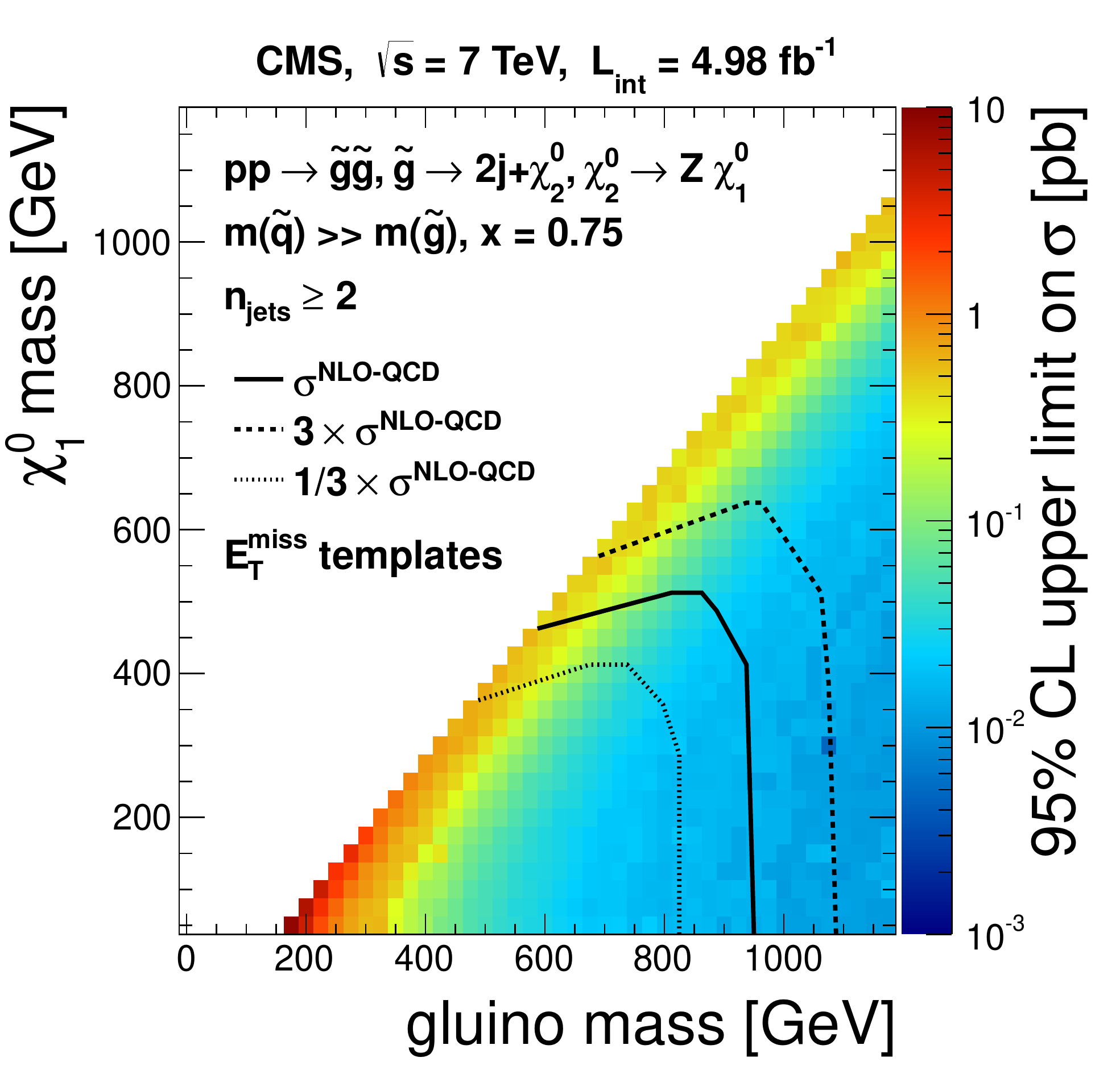}
    \caption{
\footnotesize
Limits on the SMS topology with neutralino LSP ($x=0.75$), based on the \MET\ template
method: (left) signal efficiency times acceptance normalized
to the number of events with at least one $Z\to\ell\ell$ decay for the $\MET>100\GeV$ region;
(right) 95\% CL upper limits on the total gluino pair-production cross section.
The region to the left of the solid contour is excluded assuming that the gluino pair-production
cross section is $\sigma^{\text{NLO-QCD}}$, and that the branching fraction to this SMS topology is 100\%.
The dotted and dashed contours indicate the excluded region when the cross section is varied by a factor of three.
The signal contribution to the control regions is negligible.
}
    \label{fig:T5zzl}
  \end{center}
\end{figure}

\clearpage

\begin{figure}[!ht]
  \begin{center}
    \includegraphics[width=0.48\textwidth]{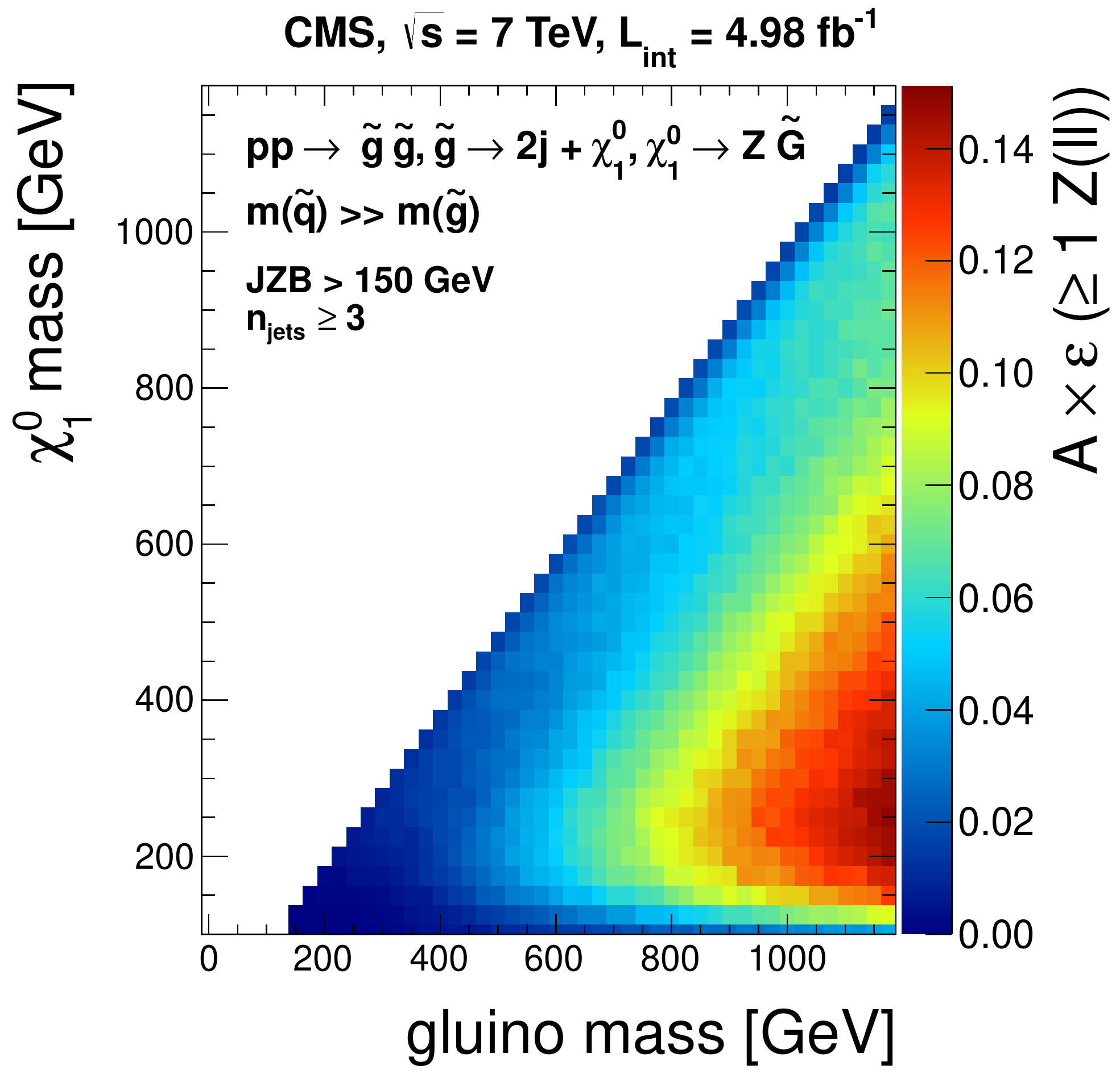}
    \includegraphics[width=0.48\textwidth]{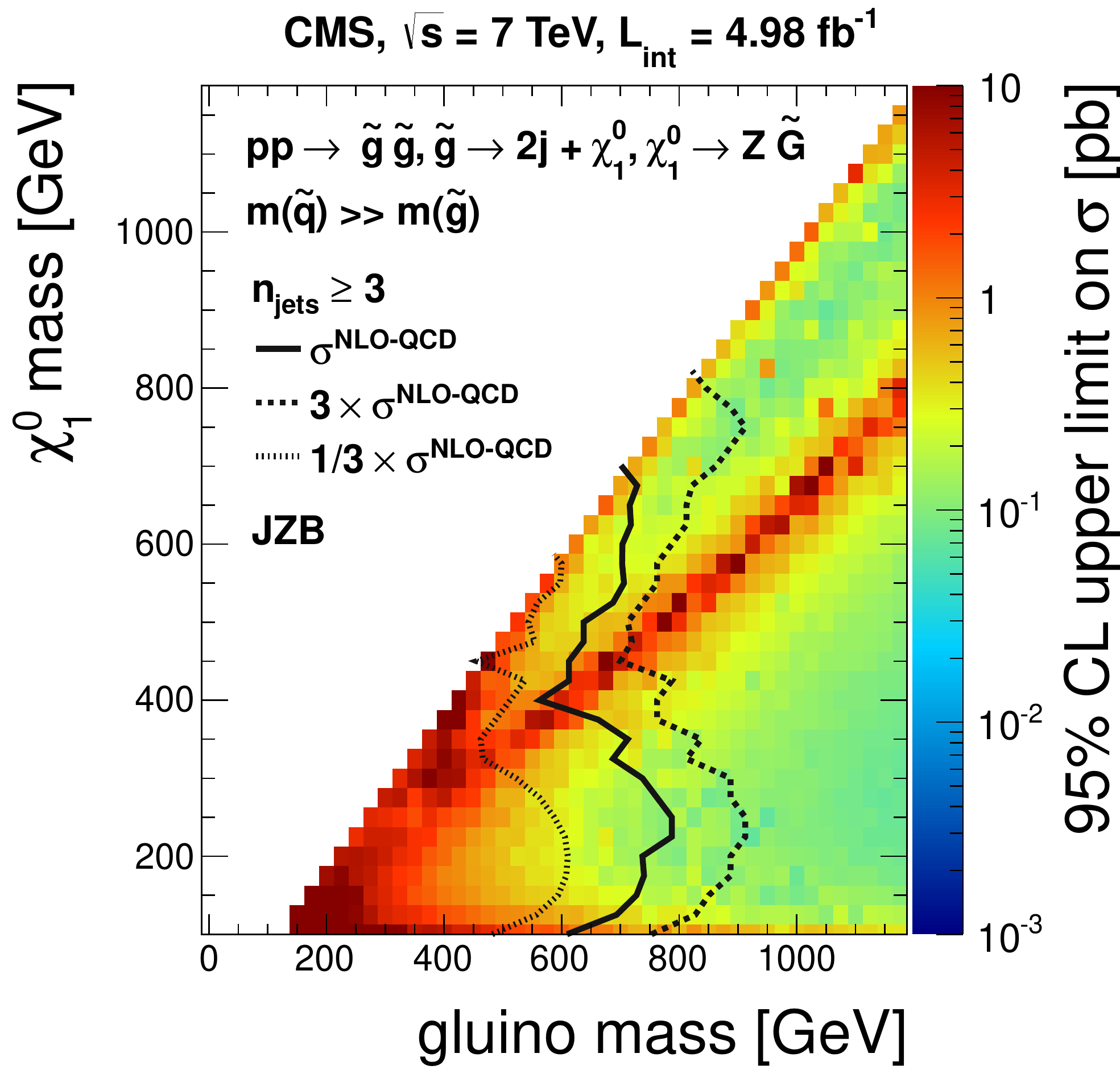}
    \caption{
\footnotesize
Limits on the SMS topology with gravitino LSP, based on the \JZB method: (left) signal efficiency times acceptance normalized
to the number of events with at least one $Z\to\ell\ell$ decay for the $\JZB>150\GeV$ region;
(right) 95\% CL upper limits on the  total gluino pair-production cross section.
The region to the left of the solid contour is excluded assuming that the gluino pair-production
cross section is $\sigma^{\text{NLO-QCD}}$, and that the branching fraction to this SMS topology is 100\%.
The dotted and dashed contours indicate the excluded region when the cross section is varied by a factor of three.
The signal contribution to the control regions is taken into account.
    \label{fig:SMS05JZB2}}
  \end{center}
\end{figure}

\begin{figure}[!hb]
  \begin{center}
    \includegraphics[width=0.96\textwidth]{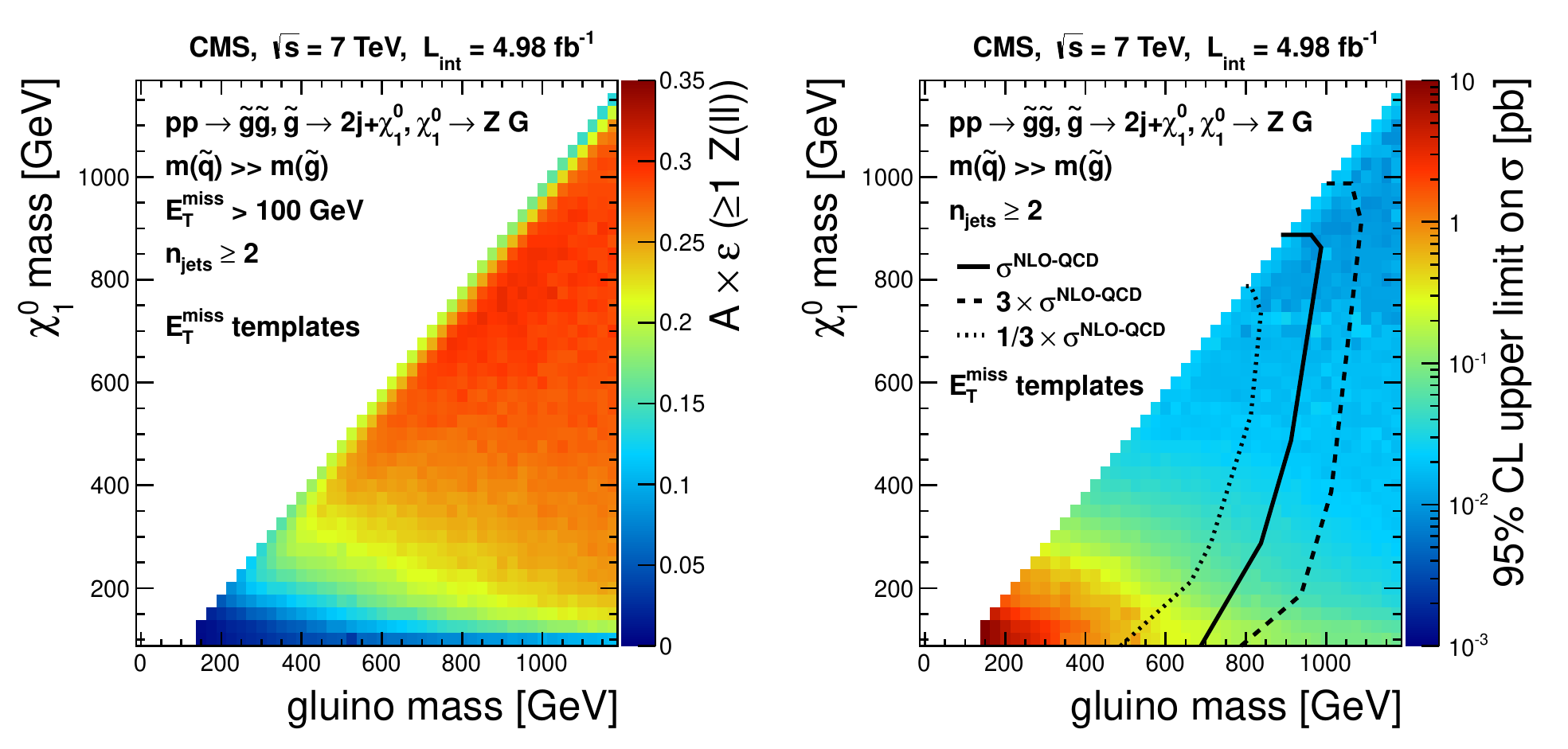}
    \caption{
\footnotesize
Limits on the SMS topology with gravitino LSP, based on the \MET\ template
method: (left) signal efficiency times acceptance normalized
to the number of events with at least one $Z\to\ell\ell$ decay for the $\MET>100\GeV$ region;
(right) 95\% CL upper limits on the total gluino pair-production cross section.
The region to the left of the solid contour is excluded assuming that the gluino pair-production
cross section is $\sigma^{\text{NLO-QCD}}$, and that the branching fraction to this SMS topology is 100\%.
The dotted and dashed contours indicate the excluded region when the cross section is varied by a factor of three.
The signal contribution to the control regions is negligible.
 }
    \label{fig:T5zzgmsb}
  \end{center}
\end{figure}

}
\cleardoublepage \section{The CMS Collaboration \label{app:collab}}\begin{sloppypar}\hyphenpenalty=5000\widowpenalty=500\clubpenalty=5000\textbf{Yerevan Physics Institute,  Yerevan,  Armenia}\\*[0pt]
S.~Chatrchyan, V.~Khachatryan, A.M.~Sirunyan, A.~Tumasyan
\vskip\cmsinstskip
\textbf{Institut f\"{u}r Hochenergiephysik der OeAW,  Wien,  Austria}\\*[0pt]
W.~Adam, T.~Bergauer, M.~Dragicevic, J.~Er\"{o}, C.~Fabjan, M.~Friedl, R.~Fr\"{u}hwirth, V.M.~Ghete, J.~Hammer\cmsAuthorMark{1}, N.~H\"{o}rmann, J.~Hrubec, M.~Jeitler, W.~Kiesenhofer, M.~Krammer, D.~Liko, I.~Mikulec, M.~Pernicka$^{\textrm{\dag}}$, B.~Rahbaran, C.~Rohringer, H.~Rohringer, R.~Sch\"{o}fbeck, J.~Strauss, A.~Taurok, F.~Teischinger, P.~Wagner, W.~Waltenberger, G.~Walzel, E.~Widl, C.-E.~Wulz
% ...
\vskip\cmsinstskip
\dag:~Deceased\\
1:~~Also at CERN, European Organization for Nuclear Research, Geneva, Switzerland\\
% ...

\end{sloppypar}
\end{document}